\documentclass[10pt,a4paper,twocolumn]{article}

\usepackage{amsmath,graphicx}
\newcommand\nc{\newcommand*}  \nc\longnc{\newcommand}

\nc\x[1]{\hskip#1em}  \nc\y[1]{\vskip#1ex}

\nc\eq[2]{\begin{align} \label{#1} #2 \end{align}}
\nc\lel[1]{\\ \label{#1}}
\nc\non{\tag*{}}
\nc\re[1]{(\ref{#1})}
\nc\m[1]{$#1$}
\nc\f\frac  \nc\ff[2]{{\textstyle \f{#1}{#2}}}
\nc\mat[1]{\begin{matrix} #1 \end{matrix}}  
\nc\smat[1]{\begin{smallmatrix} #1 \end{smallmatrix}}  

\nc\0[2]{\ifcase#1{#2}\or\lt(#2\rt)\or\lt[{#2}\rt]\or\lt\{{#2}\rt\}\or
  \mathord<{#2}\mathord>\or\lt\langle{#2}\rt\rangle\or\lt\lvert{#2}\rt
  \rvert\or\lt\lVert{#2}\rt\rVert\fi}
\nc\1[2]{\ifcase#1{#2}\or(#2)\or[#2]\or\{#2\}\or\mathord<{#2}\mathord
  >\or\langle{#2}\rangle\or\lvert{#2}\rvert\or\lVert{#2}\rVert\fi}
\nc\2[2]{\ifcase#1{#2}\or\big(#2\big)\or\big[#2\big]\or\big
  \{#2\big\}\or\big<#2\big>\or\big\langle#2\big\rangle\or\big
  \lvert#2\big\rvert\or\big\lVert#2\big\rVert\fi}
\nc\3[2]{\ifcase#1{#2}\or\Big(#2\Big)\or\Big[#2\Big]\or\Big\{#2\Big
  \}\or\Big<#2\Big>\or\Big\langle#2\Big\rangle\or\Big\lvert#2\Big
  \rvert\or\Big\lVert#2\Big\rVert\fi}
\nc\4[2]{\ifcase#1{#2}\or\bigg(#2\bigg)\or\bigg[#2\bigg]\or\bigg
  \{#2\bigg\}\or\bigg<#2\bigg>\or\bigg\langle#2\bigg\rangle\or\bigg
  \lvert#2\bigg\rvert\or\bigg\lVert#2\bigg\rVert\fi}
\nc\5[2]{\ifcase#1{#2}\or\Bigg(#2\Bigg)\or\Bigg[#2\Bigg]\or\Bigg
  \{#2\Bigg\}\or\Bigg<#2\Bigg>\or\Bigg\langle#2\Bigg\rangle\or\Bigg  
  \lvert#2\Bigg\rvert\or\Bigg\lVert#2\Bigg\rVert\fi}
\nc\9[2]{\ifcase#1{#2}\or\left(#2\right)\or\left[#2\right]\or\left
  \{#2\right\}\or\left\langle{#2}\right\rangle\or\left\langle{#2}\right
  \rangle\or\left\lvert{#2}\right\rvert\or\left\lVert{#2}\right\rVert\fi}
\nc\lt{\mathopen{}\mathclose\bgroup\left}  \nc\rt{\aftergroup\egroup\right}

\nc\Mathbf[1]{\mathchoice  
  {\hbox{\boldmath{$    #1$}}}
  {\hbox{\boldmath{$    #1$}}}
  {\hbox{\boldmath{$\scriptstyle #1$}}}
  {\hbox{\boldmath{$\scriptscriptstyle #1$}}}}

\nc\Mater[1]{\setbox1\hbox{$#1$}\copy1\kern-.91\wd1
  \widetilde{\hbox to .82\wd1{\vphantom{$#1$}}}\kern.09\wd1}
\nc\dev{^{\text{dev}}}       \nc\sph{^{\text{sph}}}
\nc\trans{^{\text{transp}}}  \nc\symm{^{\text{sym}}}
\nc\ltd[1]{{\overset{{\scriptscriptstyle \leftarrow}}{#1}}}
\nc\rtd[1]{{\overset{{\scriptscriptstyle \rightarrow}}{#1}}}
\nc\para{_{}^{\rule{.03em}{.9ex}\kern.06em\rule{.03em}{.9ex}}}
\nc\orth{_{}^{\hbox{\large$\scriptscriptstyle\perp$}}}
\nc\tr{\;\!{\text{tr}}\;\!}
\nc\dd{\mathrm{d}}  \nc\pd\partial
\nc\e{\mathrm{e}}  

\nc\Valp{\alpha}       \nc\VValp{\VE_0}
                       \nc\VVbet{\VE_1}
\nc\Vgam{\gamma}       \nc\VVgam{\VE_2}
\nc\Vdel{\delta}       \nc\VVdel{\Mathbf{\Vdel}}
\nc\Veps{\varepsilon}  \nc\VVeps{\Mathbf{\Veps}}
\nc\Vlam{\lambda}
\nc\Vpi{\pi}
\nc\Vrho{\varrho}
\nc\Vsig{\sigma}       \nc\VVsig{\Mathbf{\Vsig}}
\nc\VVtau{\tau}
\nc\Vxi{\xi}           \nc\VVxi{\Mathbf{\Vxi}}
\nc\VGam{\Gamma}

\nc\Vc{c}
\nc\Ve{e}
\nc\Vg{g}              \nc\VVg{\mathbf{\Vg}}
\nc\Vh{h}              \nc\VVh{\mathbf{\Vh}}
\nc\Vj{j}              \nc\VVj{\mathbf{\Vj}}
\nc\Vl{l}
\nc\Vp{p}
\nc\Vs{s}
\nc\Vu{u}
\nc\Vv{v}              \nc\VVv{\mathbf{\Vv}}
\nc\Vw{w}

\nc\VA{A}              \nc\VVA{\mathbf{\VA}}
\nc\VD{D}              \nc\VVD{\mathbf{\VD}}
\nc\VE{E}
\nc\VH{H}
\nc\VJ{J}              \nc\VVJ{\mathbf{\VJ}}
\nc\VL{L}              \nc\VVL{\mathbf{\VL}}
\nc\VT{T}
\nc\VZ{Z}              \nc\VVZ{\mathbf{\VZ}}

\begin{document}

\columnwidth=89.5mm  

\title{Elastic, thermal expansion, plastic and rheological processes --
theory and experiment}


\author{Csaba Asszonyi\m{^{1}}, Attila Csat\'ar\m{^{1,2}}, Tam\'as
F\"ul\"op\m{^{1,3}}}

\date{\small
\m{^1}Montavid Thermodynamic Research Group
 \\
H-1112 Budapest, Igm\'andi u. 26, Hungary
(e-mail: asszonyi@gmail.com)
 \\
\
 \\
\m{^2}Hungarian Institute of Agricultural Engineering
 \\
H-2100 G\"od\"oll\H o, Tessedik S. u. 4, Hungary
(e-mail: csatar.attila@gmgi.hu)
 \\
\
 \\
\m{^3}Department of Energy Engineering
 \\
Budapest University of Technology and Economics
 \\
H-1111 Budapest, Bertalan L. u. 4-6, Hungary
(e-mail: fulop@energia.bme.hu)
 \\
\
}

\maketitle

\columnwidth=89.5mm  

 \begin{abstract}
  \it
Rocks are important examples for solid materials where, in various
engineering situations, elastic, thermal expansion,
rheological/viscoelastic and plastic phenomena each may play a
remarkable role. Nonequilibrium continuum thermodynamics provides a
consistent way to describe all these aspects in a unified framework.
This we present here in a formulation where the kinematic quantities
allow arbitrary nonzero initial (e.g., in situ) stresses and such
initial configurations which -- as a consequence of thermal or remanent
stresses -- do not satisfy the kinematic compatibility condition. The
various characteristic effects accounted by the obtained theory are
illustrated via experimental results where loaded solid samples undergo
elastic, thermal expansion and plastic deformation and exhibit
rheological behaviour. From the experimental data, the rheological
coefficients are determined, and the measured temperature changes are
also explained by the theory.

{Dedicated to the memory of Zolt\'an Szarka (1927--2015).}
 \end{abstract}

\vskip2ex\noindent\textit{\textbf{Keywords}}
 \par\noindent{\it
solids, elasticity, rheology, thermal expansion, plasticity,
thermodynamics}

\section{Introduction}

Motivated by problems in rock mechanics and similar challenges in the
continuum theory of solids, in the last few years, our research object
has been to achieve an amalgamation of a new approach
\cite{FulVan10,Ful11,FulVanCsa13Bre,FulVanCsa13Bal} to the problem of
objectivity and of material frame indifference -- based on Matolcsi's
framework \cite{Mat84,Mat93,Mat86,MatGru96} -- with a recent
activity in nonequilibrium thermodynamics
\cite{Van96,VanVas01,VanFul12,AssFulVan14} that focuses on the role of
thermodynamical stability and on a constructive quantitative
exploitation of the content of the second law of thermodynamics. Here,
we present how this program has accomplished a theoretical framework for
the continuum thermo-elasto-visco-plasto-mechanics of solids.

Accordingly, the aspects covered currently are:

 \begin{list}{\m{\bullet}}{\partopsep=0ex\topsep=0ex\itemsep=0ex\parsep=0ex}
 \item
elasticity, an immediate response to mechanical loading, and during
which mechanical energy is conserved;
 \item
rheology, which, in contrast, is a delayed response, with mechanical
energy partially dissipated, and which may be attributed to viscous
damping, for instance;
 \item
plasticity, which is permanent shape change caused by mechanical
loading: a change of the unloaded shape;
 \item
thermal expansion, and the thermal stress generated by it;
 \item
and heat conduction.
 \end{list}
In parallel to the general level -- large deformation theory, general
constitutive equations --, we consider it inevitable to exhibit (and
countercheck!) the applicability of the formulation to practical
concrete examples. To this end, we have performed experiments on which
the theory can be applied and tested. The experimental results presented
here illustrate the various predictions of the theory both qualitatively
and quantitatively. Via this simple yet widely informative and
insightful experimental example -- mechanical and thermal monitoring of
uniaxial stretching of polyamide-6 plastic samples --, the various
thermomechanical effects in solids are well demonstrated and the
correspondence to the theoretical expectations are satisfactorily
established.

Therefore, though the theoretical framework described here is capable to
describe completely general situations, here our aim is to focus on the
connection between theory and experiment so, at each component of our
theoretical formulation, we take the simplest applicable concrete
choice: small deformations, Hooke elasticity with constant elastic
coefficients, constant specific heat, thermal expansion and heat
conduction coefficients, and constant plastic change rate coefficient
and yield stress. These assumptions satisfactorily suit the obtained
experimental data.

We start the discussion with the kinematic quantities according to the
mentioned recent objective approach. Then we build elasticity, thermal
expansion and plasticity around these quantities, via continuum
thermodynamics. We show how the predicted behaviours can be observed
experimentally. Next, we incorporate rheology into the theory, and point
out how the example measurement results indicate the presence of
rheology both on the mechanical and the thermal side. We determine the
rheological coefficients from the experimental data, and discuss the
related arising numerical challenges and the used solutions that enabled
us to obtain these coefficients reliably. In the Discussion section, we
summarize the most important outcomes and lessons, and outline the
future perspectives and aims. Technical details concerning the
measurement can be found in the Appendix.

\section{The kinematic ingredients}\label{vac}

We start with a succint account of the objectivity respecting definition
of the involved kinematic quantities, presented in detail in
\cite{FulVan10,Ful11,FulVanCsa13Bre,FulVanCsa13Bal}. For continua --
solids, liquids and gases each -- the motion determines the
instantaneous distance of any two material points. This defines an
\textit{instantaneous metric} \m{ \Mater\VVh } on the material manifold.
Hereafter, overtilde indicates tensors on the material manifold,
operating on the material tangent vectors. For solids, the novel and
important notion is the \textit{relaxed or self-metric} \m{ \Mater\VVg
}, which describes the distances of material points when the solid is in
an unstressed, relaxed state. In such a state, \m{ \Mater\VVh =
\Mater\VVg }. The relaxed metric characterizes the relaxed shape of a
solid body.

For the purposes of the elastic state variable, the appropriate
kinematic quantity can be defined from
 \eq{fcj}{
\Mater\VVA = \Mater\VVg^{-1} \Mater\VVh ,
 }
the \textit{elastic shape} symmetric tensor -- the (objective
generalization of the) right Cauchy--Green tensor --, as
 \eq{@12367}{
\Mater\VVD = \f{1}{2} \ln \Mater\VVA.
 }
This \textit{elastic deformedness} tensor \m { \Mater\VVD } is on which
elastic stress, \m{ \Mater\VVsig }, is considered to depend on, linearly
(Hooke's law) or nonlinearly (e.g., a neo-Hooke model). This logarithmic
type -- objectively generalized Hencky strain -- definition \re{@12367}
is distinguished both geometrically (the spherical/trace part of
describes the volumetric change and the deviatoric part corresponds to
the constant-volume changes, even for large deformations
\cite{FulVan10,NefEta14,NefEta15}) and experimentally (for example,
large-deformation stress in materials like hard rubber is most linear in
this logarithmic tensor \cite{BeaSta86,HorMur09}, the five-parameter
Murnaghan model of nonlinear elasticity also performs the best in this
logarithmic variable \cite{PleKru06}, and see also \cite{BruEta01}.

Via the (objective generalization of the) deformation gradient, the
material tangent vectors can be mapped to the spatial vectors. Our
material tensors (\m{ \Mater\VVh }, \m{ \Mater\VVA } etc.) are
correspondingly mapped to their spatial counterpart (\m{ \VVh }, \m{
\VVA } etc.). The change of \m{ \Mater\VVh } in time, and thus the time
derivative of \m{ \VVA }, can be calculated with the aid of the velocity
gradient
 \eq{@12981}{
\VVL = \VVv \otimes \ltd{\nabla}
 }
(and its transpose \m { \VVL\trans = \rtd{\nabla} \otimes \VVv}), and
one finds
 \eq{@13604}{
\dot\VVA = \VVL \VVA + \VVA \VVL\trans
 }
for the comoving time derivative \m { \dot\VVA }, as long as only
elasticity is involved (\m { \Mater\VVg = \text{const.} }).

Thermal expansion is the phenomenon that the unstressed and relaxed
size, hence, the relaxed metric, of solid bodies depends on temperature:
\m { \Mater\VVg = \Mater\VVg (\VT) }(\m { \VT } denoting absolute
temperature throughout this paper). If the material is isotropic, which
we assume in what follows, then this temperature dependence is a simple
scalar rescaling:
 \eq{fcl}{
\Mater\VVg{\1 1{\VT_2}} & = \Lambda{\1 1{\VT_1, \VT_2}}^2
\Mater\VVg{\1 1{\VT_1}},
 \lel{fcn}
\Valp \1 1 {\VT} & = \lim\limits_{\Delta \VT \to 0}^{} \f{ \Lambda \9 1
{ \VT, \VT + \Delta \VT } - 1 }{ \Delta \VT } ,
 }
the latter formula defining the linear thermal expansion coefficient
\m{\Valp}. It follows that, when temperature changes in time at a
material point, we have
 \eq{fbq}{
\dot{\Mater{\VVg}} = \9 1{ \f{\dd}{\dd \VT} \Mater{\VVg} } \dot \VT = 
2 \Valp(\VT) \dot \VT \, \Mater{\VVg}
 }
for the comoving time derivative of the relaxed metric, 
and \re{@13604} is generalized to
 \eq{@13605}{
\dot\VVA = \VVL \VVA + \VVA \VVL\trans - 2 \Valp \dot\VT \VVA .
 }
Plasticity (see, e.g., the recent monography \cite{RusRus11}) is, in our
language, another phenomenon involving the change of \m { \Mater{\VVg}
}: strong enough mechanical stress causes the relaxed shape -- and
metric -- of a solid to change permanently. This change can be
characterized by the \textit{plastic change rate} tensor
 \eq{@15312}{
\Mater\VVZ = \f {1}{2} \Mater{\VVg}^{-1}
\dot{\Mater{\VVg}}_{\text{plastic}} ,
 }
with which the total kinematic time evolution equation is
 \eq{@13606}{
\dot\VVA = \VVL \VVA + \VVA \VVL\trans - 2 \Valp \dot\VT \VVA - 2 \VVZ
\VVA .
 }
In the subsequent sections, we restrict ourselves to the
small-deformation regime, \m { \1 7 {\VVD} \ll 1 }, \m { \VVA = \e^{2
\VVD} \approx \mathbf{1} + 2 \VVD }, where \re{@13606} leads, in the
leading order of \m { \VVD }, to \m {2 \dot\VVD = \VVL + \VVL\trans - 2
\Valp \dot\VT \mathbf{1} - 2 \VVZ }, rearrangable as
 \eq{@15991}{
\VVL\symm = \dot\VVD + \Valp \dot\VT \mathbf{1} + \VVZ
 }
(\m { \symm } standing for symmetric part). Further, for our purposes
below, the thermal expansion coefficient can be taken as constant.

\section{Comparison with the usual approach to kinematic quantities}

Conventionally, a reference frame and an initial time \m { t_0 } is
chosen, and displacements are considered in the space of this reference
frame and are measured related to positions at \m { t_0 }. From the
displacements, the deformation gradient is defined, and the various
strain measures are constructed from the deformation gradient. One
consequence of this approach is that initial strains are necessarily
zero, and the corresponding elastic stress is also zero. Initial strains
also necessarily satisfy the compatibility condition.

Furthermore, when thermal expansion is also considered then a
homogeneous initial  temperature distribution is assumed as well.

One of our aims with the alternative kinematic formulation expounded in
the preceding section was to generalize this approach to situations
where initial (e.g., in situ) stresses are unavoidably nonzero, the
initial temperature distribution is far from homogeneous, or where in situ
or remanent stresses indicate that the compatibility condition is
violated already initially. Namely, as we have shown
\cite{FulVan10,Ful11,FulVanCsa13Bre,FulVanCsa13Bal}, while the
instantaneous metric \m { \Mater\VVh } is a flat Riemann metric by
definition, the relaxed Riemann metric \m { \Mater\VVg } is not
necessarily flat (so the compatibility condition -- see its finite
deformation version lengthy formula in \cite{FulVan10,Ful11} -- is
violated), for example as a consequence of some plastic preceding
history or an inhomogeneous temperature distribution. Then these two
metrics necessarily differ at some parts of the material, causing there
nonzero elastic deformedness, and thus leading to nonzero elastic stress
(remanent, "frozen" stress).

In our generalized formulation, strains -- which are nevertheless
important notions for experimental purposes -- can be given as time
integrals of the various terms of \re{@15991} from reference time to
current time: total strain \m { \VVeps } is the time integral of the lhs
term, elastic strain \m { \VVeps_{\text{el}} } is that of the first term
on the rhs, thermal expansion strain \m { \VVeps_{\text{th}} } is that
of the second one, and plastic strain \m { \VVeps_{\text{pl}} } is the
integral of the third term. In the small deformation regime involved in
the experiments below, these definitions suffice and no finite
deformation complication needs to be addressed.

\section{Mechanics and thermodynamics}\label{fbh}

Our chosen elastic constitutive equation is Hooke's law:
 \eq{fbw}{
\VVsig = \VE\dev \VVD\dev + \VE\sph \VVD\sph
 }
with the spherical and deviatoric components and elastic coefficients
 \eq{fbx}{
\VVD\sph & = \ff{1}{3} (\tr \VVD) \mathbf{1},
 &
\VVD\dev & = \VVD - \VVD\sph ,
 \lel{fco}
\VE\sph & = 3K ,
 &
\VE\dev & = 2G .
 }
For us here, it suffices to consider \m { \VE\dev } and \m { \VE\sph }
as constant. From \re{fbw}, it is easy to show \cite{Ful11} that the
classic Duhamel--Neumann formula for thermoelasticity \cite{Lub04} is
recovered as a special case, via transforming from the elastic
deformedness variable to the total strain and imposing that, at the
initial reference time,
 \begin{list}{\m{\bullet}}{\partopsep=0ex\topsep=0ex\itemsep=0ex\parsep=0ex}
 \item
elastic deformedness is zero,
 \item
the temperature distribution is homogeneous,
 \item
and plastic change does not occur.
 \end{list}

The first law of continuum thermodynamics, i.e., the balance of internal
energy, reads
 \eq{fbz}{
\Vrho \dot \Ve = - \VVj_{\Ve} \cdot \ltd\nabla + \tr \9 1 { \VVsig
\VVL\symm }
 }
for the specific internal energy \m { \Ve \1 1 { \VVD, \VT } } and its
current \m { \VVj_{\Ve} }, both to be specified constitutively (and the
mass density \m { \Vrho } being constant in the small deformation
regime). Similarly, the balance for specific entropy \m { \Vs  \1 1 {
\VVD, \VT } } is
 \eq{fca}{
\Vrho \dot \Vs = { - \VVj_{\Vs} \cdot \ltd\nabla + \Vpi_{\Vs} } ,
 }
where we take the simplest and standard choice \m { \VVj_{\Vs} =
\VVj_{\Ve}/\VT } for the the entropy current \m { \VVj_{\Vs} }, \m { \Vs
} must be thermodynamically consistent with \m { \Ve } (i.e., the Gibbs
relation must hold between them), and entropy production must be
positive definite, \m { \Vpi_{\Vs} \ge 0 }. Specifically, we take
 \eq{fcf}{
\VVj_{\Ve} = \Vlam \rtd\nabla \f{1}{\VT} ,
 }
that is, standard Fourier heat conduction, with positive coefficient \m
{ \Vlam }, and
 \eq{fcb}{
\Ve & = \Vc \VT + \9 3 { \f{\VE\dev}{2\Vrho} \tr \0 2 { \9 1 { \VVD\dev
}^2 } + \f{\VE\sph}{2\Vrho} \tr \0 2 { \9 1{ \VVD\sph }^2 } }
 \non\lel{fcp}
& \quad
+ \f{\VE\sph}{\Vrho} \VT \Valp \tr \VVD\sph ,
 }
for specific internal energy, in which the first term provides the
constant specific heat \m { \Vc }, the middle term describes elastic
energy, and the last one is responsible for thermal expansion. The
corresponding specific entropy is determined up to an additive constant,
as
 \eq{fcd}{
\Vs = \Vc \ln \f{\VT}{\VT_*} + \f{\VE\sph}{\Vrho} \Valp \tr \VVD\sph
 }
with an arbitrarily auxiliary constant \m { \VT_* }. In the resulting
entropy production,
 \eq{fce}{
\Vpi_{\Vs} & = \rtd\nabla \f{1}{\VT} \cdot \VVj_{\Ve} + \f{1}{\VT} \tr
\0 1 { \VVsig \VVZ }
 \lel{fcq}\non
& = \rtd\nabla \f{1}{\VT} \cdot \VVj_{\Ve}
+
\f{\VE\dev}{\VT} \tr \1 1 { \VVD\dev \VVZ }
 + \f{\VE\sph}{\VT} \tr \1 1 { \VVD\sph \VVZ } .
 }
the first term of the rhs is non-negative due to \re{fcf}, and we ensure the
two further terms to be positive definite by choosing the plastic
constitutive equation as
 \eq{fcg}{
\VVZ = \VGam \dot\VVD\dev
 }
with
 \eq{fcr}{
\VGam & = \Vgam \VH \0 1{ \tr \0 2 { \9 1{ \VVD\dev }^2 } - \f {2}{3}
\VD_{\text{crit}}^2 }
 \non\lel{fcs}
& \quad \times \VH \0 1{ \tr \0 2 { \9 1{ \VVD\dev \dot\VVD\dev } } } ,
 }
where \m { \Vgam } and \m { \VD_{\text{crit}} } are positive constants,
and \m { \VH } is the Heaviside function.

Equation \re{fcg} describes a natural plasticity theory:
 \begin{list}{\m{\bullet}}{\partopsep=0ex\topsep=0ex\itemsep=0ex\parsep=0ex}
 \item
the plastic change rate is proportional to the deviatoric elastic change
rate,
 \item
the yield criterion is the von Mises one (note that, now, stress is in a
Hookean relationship with elastic deformedness so \m { \VVD } can be
replaced by \m { \VVsig }, and \m { \VD_{\text{crit}} } is equivalent to
a \m { \Vsig_{\text{crit}} }, the von Mises yield stress),
 \item
and plastic change is switched off during
unloading, as a consequence of the second Heaviside factor that ensures
the thermodynamical requirement \m { \Vpi_{\Vs} \ge 0 }.
 \end{list}

For temperature dependent coefficients \m { \VE\sph }, \m { \VE\dev },
\m { \Valp }, \m { \Vc }, for more general internal energy, for large
deformations, and for anisotropic materials, one can derive similar
though somewhat more complicated formulae.

The change of temperature is determined by the rate equation derivable
from \re{fcp} with \re{fbz}, \re{fbw} and \re{@15991} substituted:
 \eq{@25278}{
\Vrho \Vc \dot \VT = - \VVj_{\Ve} \cdot \ltd\nabla - \VE\sph\Valp \VT
\tr \dot\VVD\sph + \VE\dev \tr \1 1 { \VVD\dev \VVZ } .
 }
The first term here gives account of the effect of heat. The second term
is the source of the Joule--Thomson effect for solids: cooling during
stretching and warming during compression. This is a reversible type
change, a less obvious but inevitable manifestation of thermal
expansion. The third term, on the contrary, describes an irreversible
effect, being non-negative -- as a result of the non-negativity of
entropy production -- and thus always causing warming whenever plastic
change takes place.

Adding the mechanical equation of motion,
 \eq{@25967}{
\Vrho \dot \VVv = \VVsig \cdot \ltd\nabla ,
 }
or its force-equilibrial approximation (applied in what follows)
 \eq{@25968}{
\VVsig \cdot \ltd\nabla = \mathbf{0} ,
 }
to the balance and constitutive equations above, we arrive at a closed
set of dynamical equations: \re{@12981}, \re{@15991}--\re{fbw}, \re{fcf},
\re{fcg}--
\re{@25278} and \re{@25967} \1 2 {or
\re{@25968}}. \1 1 {Volumetric forces can, naturally, be added if
necessary.} Therefore, we can calculate any concrete process, provided
the required amount of initial and boundary conditions are at hand.

\section{Uniaxial processes -- formulae and experimental illustration}
\label{fdv}

We demonstrate the application of the obtained theoretical framework on
uniaxial processes. Such situations are seminal because of two reasons:
they are simple to calculate, and are capable to describe many
experimental tests. For bodies with such special geometry, and assuming
adiabaticity as well as symmetry respecting space independent boundary
conditions, all quantities have a homogeneous distribution. In other
words, quantities are time dependent but space independent. In
appropriate coordinate system, tensors have at most only longitudinal
(\m { || }) and transversal (\m { \perp }) components:
\nc\Vnull{\vphantom{|^|}\hphantom{|}0\hphantom{|}}
 \eq{.17.62.}{
\VVsig & = \0 1{ \smat{ \Vsig\para & & \\ & \Vnull & \\ & &
\Vnull } } ,
 &
\VVD & = \0 1 { \smat{ \VD\para & & \\ & \VD\orth & \\ & & \VD\orth } } ,
 \non\lel{.17.63.} 
\VVeps & = \0 1 { \smat{ \Veps\para & & \\ & \Veps\orth & \\ & & \Veps\orth } }
 &
& \text{etc.}
 }
The deviatoric and spherical parts read, consequently - shown on the
example of \m { \VVD }:
 \eq{@27697}{
\VVD\dev & = \f{1}{3} \0 1{ \smat{ 2 \0 1{ \VD\para - \VD\orth }
& & \\ & - \0 1{ \VD\para - \VD\orth } & \\ & &
- \0 1{ \VD\para - \VD\orth } } } ,
 \non\lel{fct}
\VVD\sph & = \f{1}{3} \0 1{ \smat{ \VD\para + 2 \VD\orth & & \\
& \VD\para + 2 \VD\orth & \\ & & \VD\para + 2 \VD\orth } } .
 }

Let us consider a mechanical equilibrial initial condition: \m { \VVD\1
1 { t_0 } = \mathbf{0} } at an initial time \m { t_0 } -- when,
naturally, no plastic change takes place: \m { \VVZ\1 1 { t_0 } =
\mathbf{0} }, and the purely elastic stress is zero --, and let \m {
\VT_0 } denote the initial temperature. The total strain \m { \VVeps }
-- measured in experiments directly --, is, as mentioned earlier, the
time integral of \m { \VVL\symm } counted from \m { t_0 } so it also
starts from zero at initial time. From a finite difference numerical
perspective, the solution can be determined as follows. For
definiteness, let us consider the case of force-driven process, where \m
{ \Vsig\para \1 1 { t } } is prescribed (taking into consideration that,
in the small-deformation regime, the change of cross-section can be
neglected). For strain-driven processes, a similar but somewhat more
refined scheme is needed. Notably, the finite difference scheme outlined
here is, despite its simplicity, suitable to demonstrate that the
solution of the problem is well-defined. In addition, for many solid
mechanical applications, its preciseness and performance may suffice
with moderately small time steps.

So, assuming that all quantities are known up to time \m { t }, from the
prescribed \m { \Vsig \para \1 1 { t + \Delta t } }, which actually
means the knowledge of \m { \VVsig \1 1 { t + \Delta t } }, we determine
\m { \VVD\1 1 { t + \Delta t } } via \re{fbw}. Next, using \m { \1 2 { 
\VVD\1 1 { t + \Delta t } - \VVD\1 1 { t } } / \Delta t } as an
approximation for \m { \dot\VVD } during the interval \m { \1 2 { t, t +
\Delta t } }, we can apply \re{fcg} to calculate \m { \VVZ } for this
interval. Subsequently, \re{@25278} leads to a prediction of \m {
\dot\VT }, which then offers \m { \VT \1 1 { t + \Delta t } }. Also, it
enables to determine \m { \VVL\symm = \dot\VVeps } from \re{@15991} so,
at last, we obtain \m { \VVeps \1 1 { t + \Delta t } } (which data is
useful for comparison with experimentally measured values).

\begin{figure}[ht]
\parbox{77.5 mm}{%
 \begin{center}
 \includegraphics[width=.5\columnwidth]{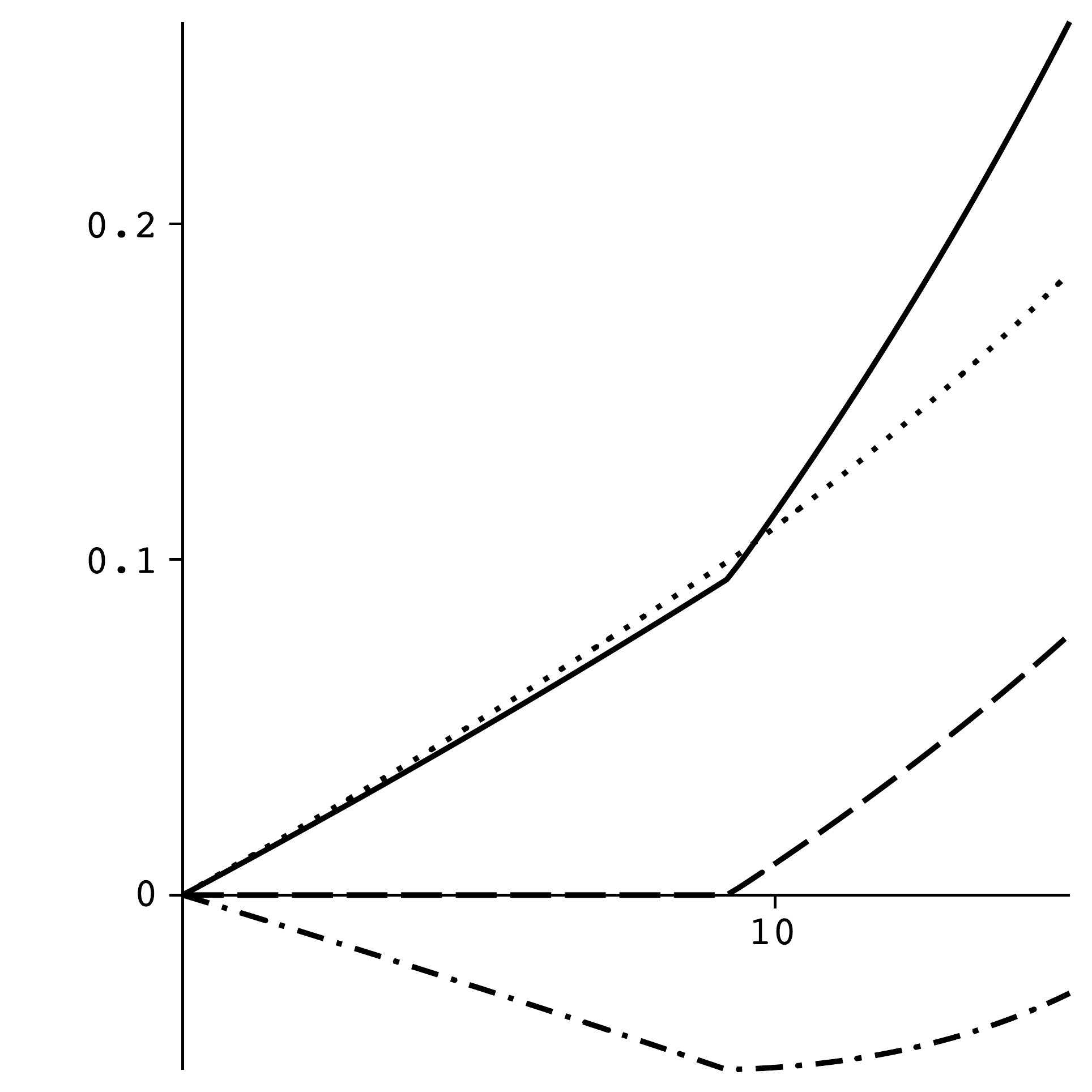}
\caption{Characteristic time dependence of strains and temperature
during uniaxial stretching by increasing force, according to the theory
\1 1 {arbitrary units, semi-quantitative plot: The thermal expansion
coefficient is set unrealistically high to make its contribution
visible}. Temperature (dash-dotted line) first decreases -- like for an
adiabatically expanding gas -- and then increases -- due to plastic
dissipation --, elastic strain (dotted line) increases as it follows the
increased stress, and plastic strain (dashed line) appears only above
the critical stress, causing that the total strain (solid line) starts
to increase faster.}%
 \label{fbf}
\end{center}
}
\end{figure}

If a sample starts from the considered initial conditions, i.e., a
relaxed and equilibrial state, and undergoes stretching with increasing
force, the obtained theory predicts the following qualitative behaviour
(see Fig.~\ref{fbf}). Below the yield threshold, the Joule--Thomson
effect is observable, and decreasing temperature results in some thermal
contraction, which acts against the increase of elastic deformedness
(but the latter dominates). When plastic change enters, it adds to the
total strain increase, and the corresponding dissipation is a
temperature increasing effect.

\begin{figure}[ht]
\parbox{77.5 mm}{%
\begin{center}
 \includegraphics[width=.57\columnwidth]{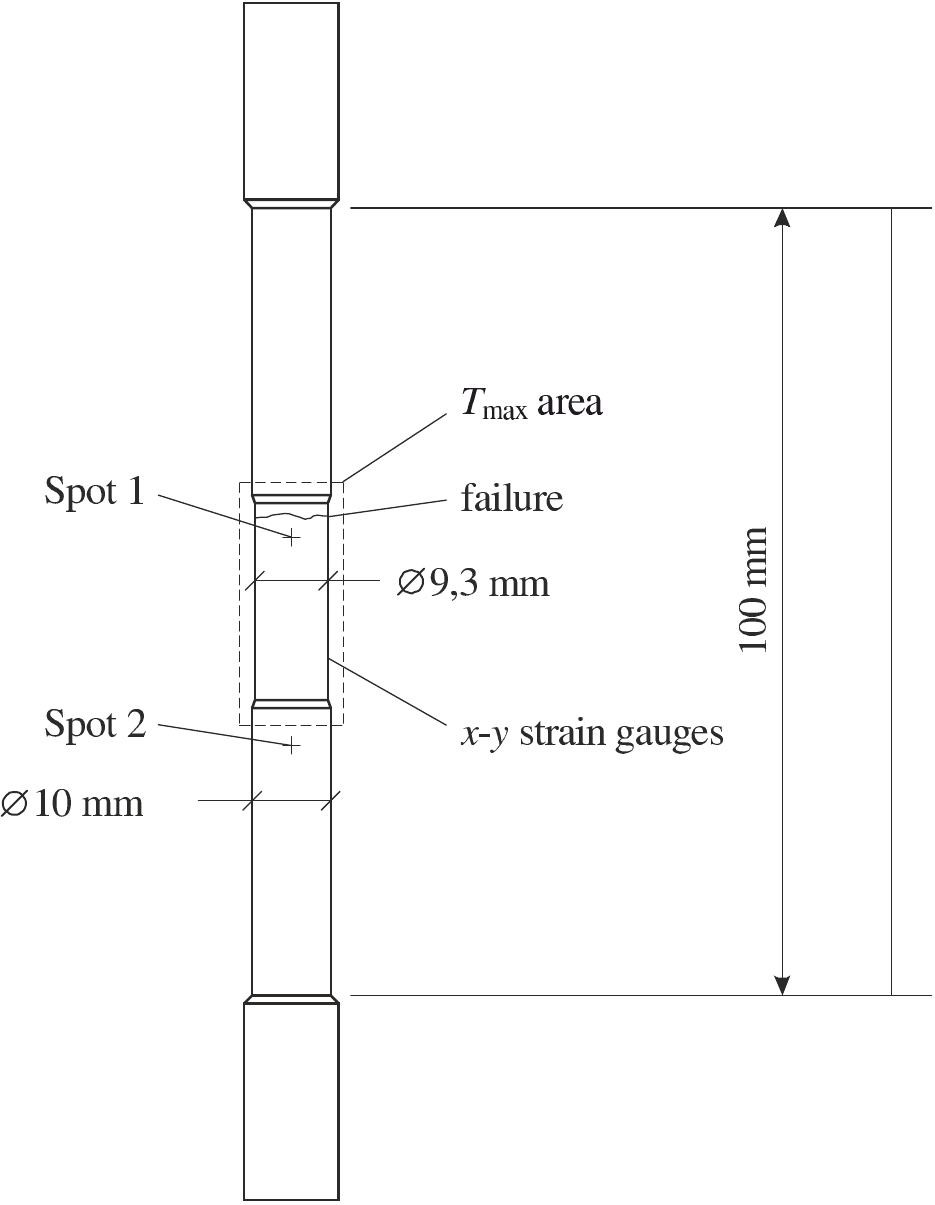}
\caption{The outline of the experiment. The middle part of the sample
was thinner, and was monitored by a thermal camera. Temperature values
at the two displayed spots were numerically displayed (see
Fig.~\ref{fbv}), together with the maximal temperature in the rectangle
area.}
 \label{fbu}
\end{center}
}
\end{figure}

These phases can nicely be demonstrated experimentally when one monitors
the temperature aspect during the stretching process. Here, we show five
snapshots taken by a thermal camera during an example experiment
performed on a polyamide-6 plastic sample (see Fig.~\ref{fbu}), which
depict the important stages (see Fig.~\ref{fbv}). The last stage,
failure, is not modelled here theoretically but is a phenomenon also
expected to be possible to incorporate in a thermodynamical formulation
\cite{Van96,VanVas01}. Namely, failure is likely to be a loss of
thermodynamical stability of the continuum.

\newlength{\tffigw}  \setlength{\tffigw}{.28\columnwidth}
\begin{figure}[ht]
\parbox{77.5 mm}{%
\begin{center}
 \includegraphics[width=\tffigw]{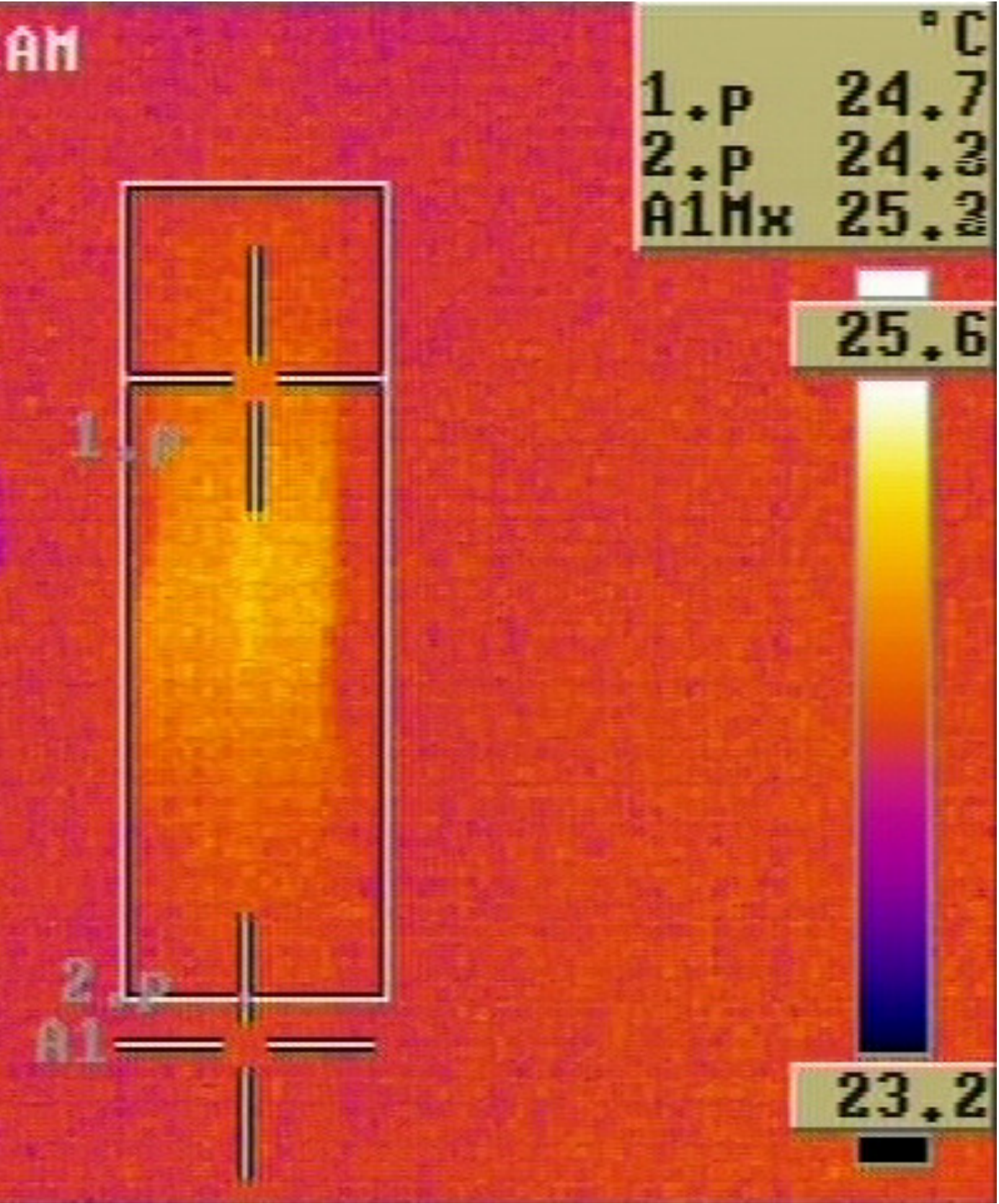}
 \hfill
 \includegraphics[width=\tffigw]{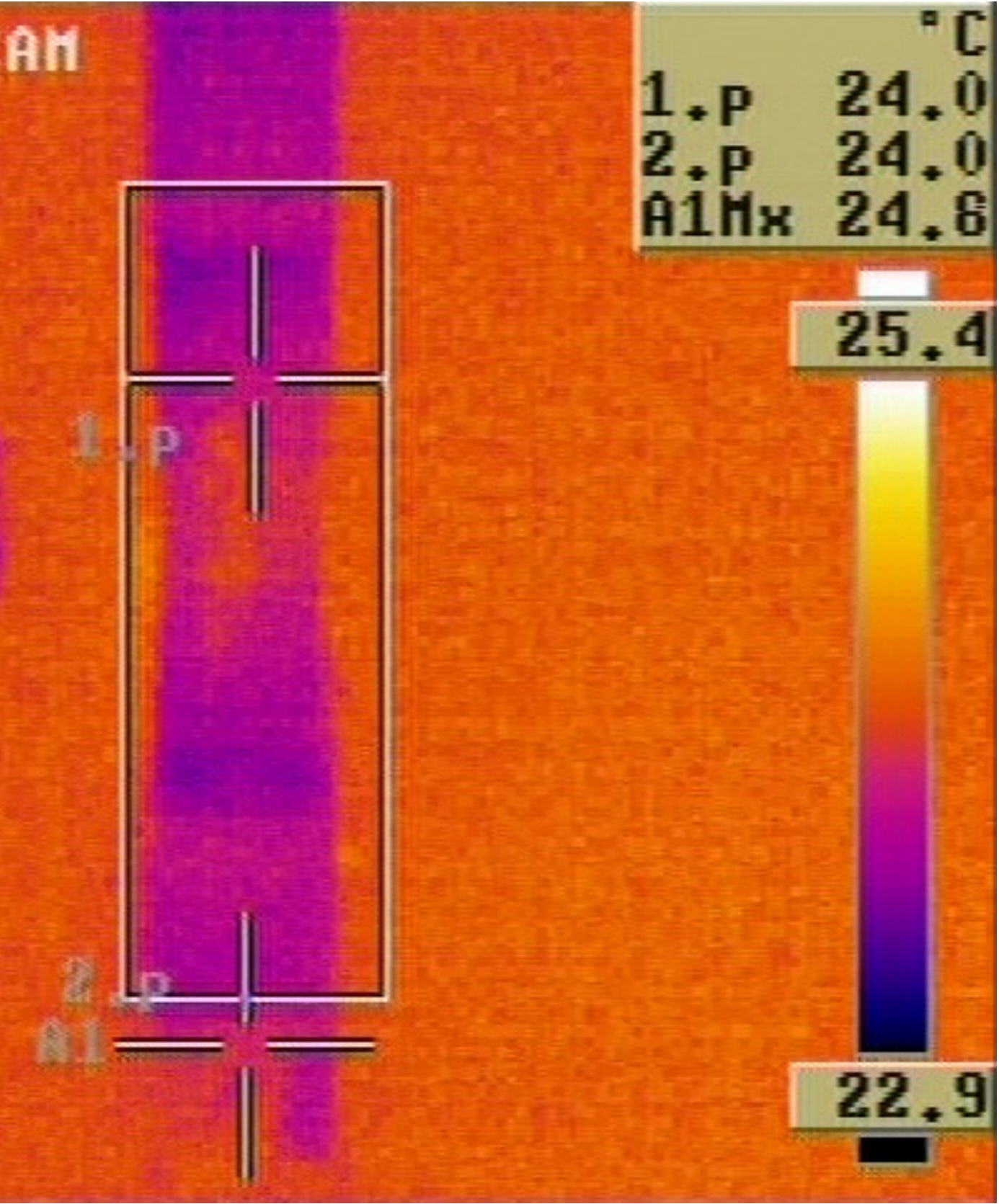}
 \hfill
 \includegraphics[width=\tffigw]{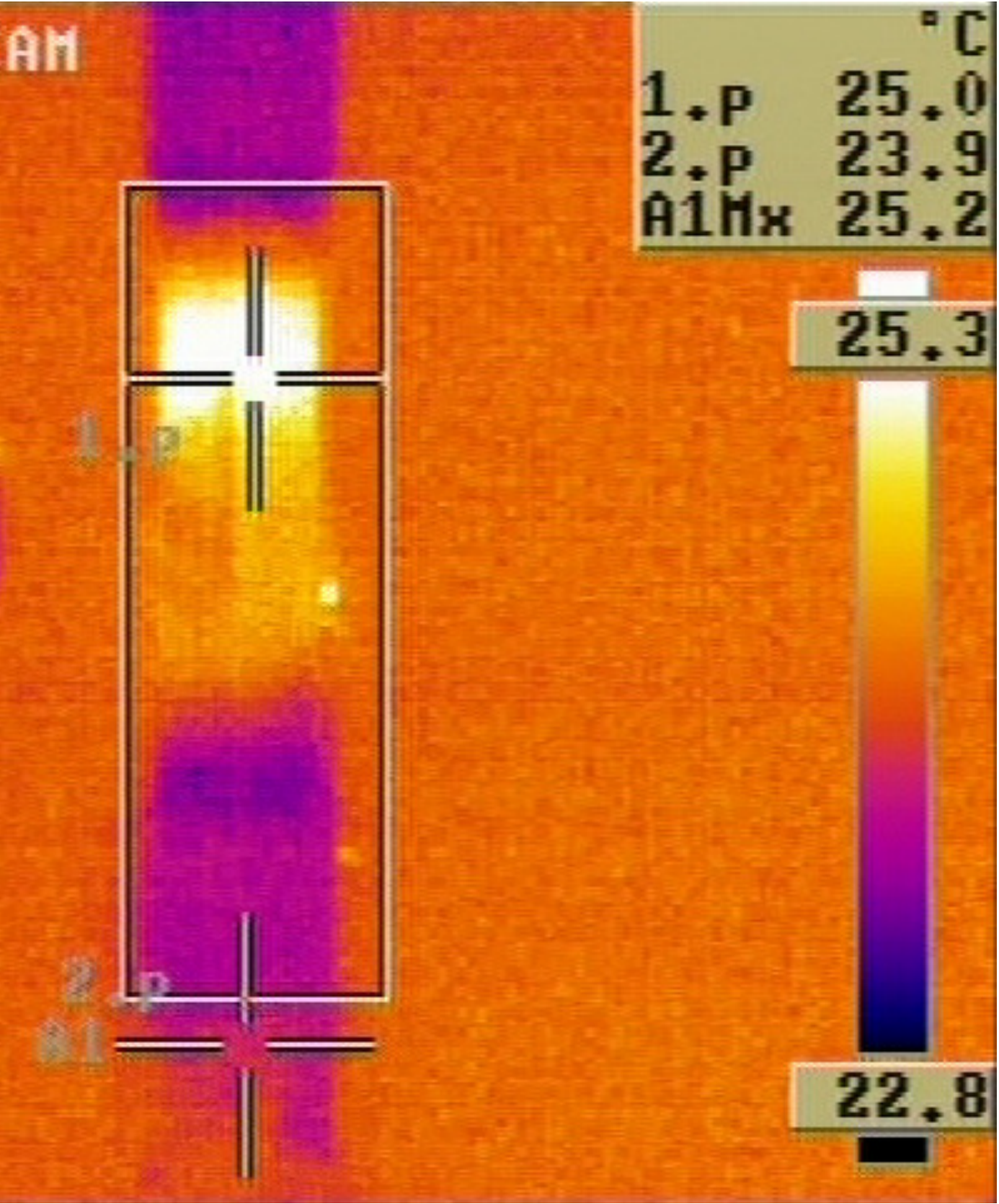}
\par\noindent
\null
 \hfill
 \includegraphics[width=\tffigw]{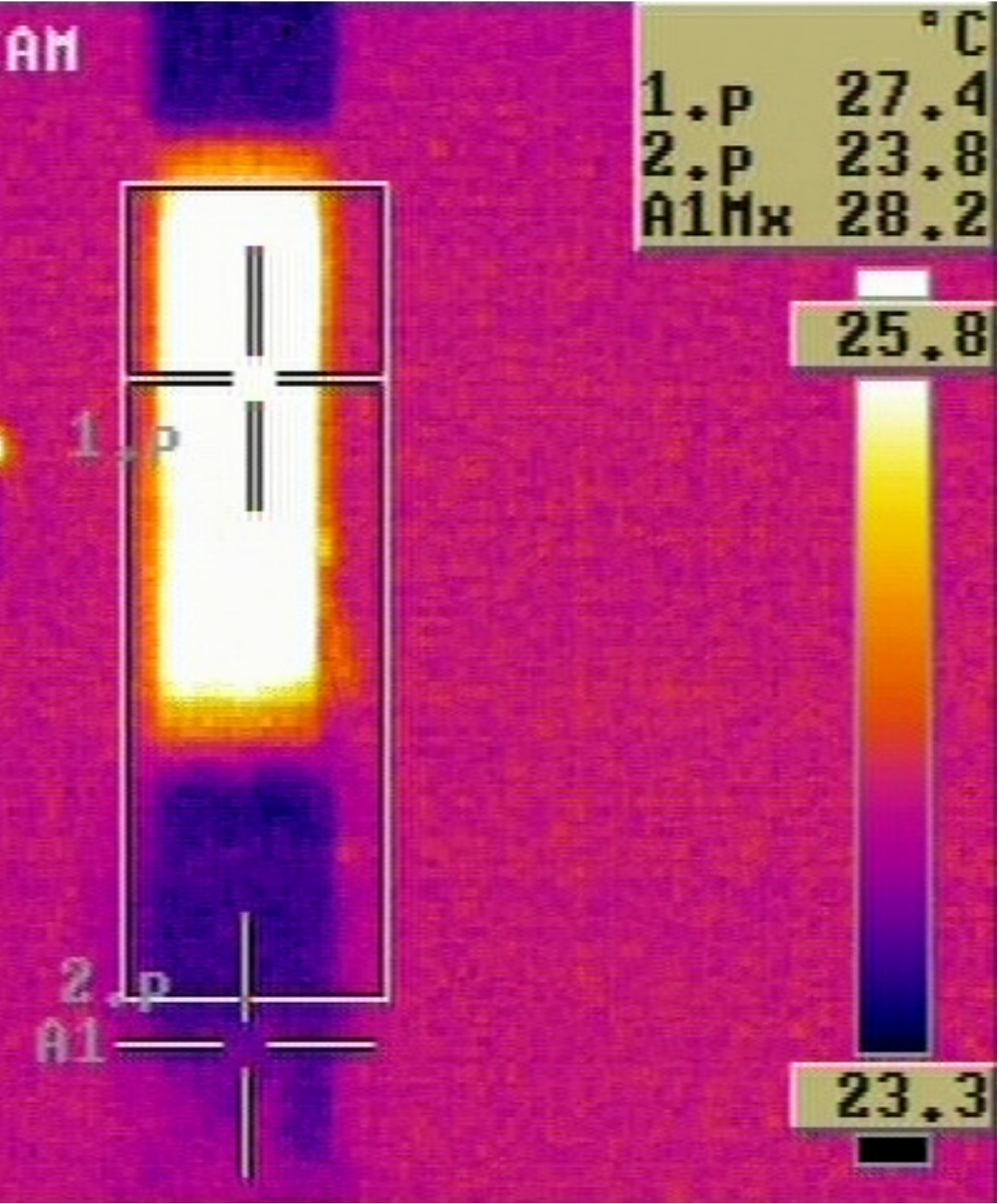}%
 \rule{.015\columnwidth}{0ex}%
 \includegraphics[width=\tffigw]{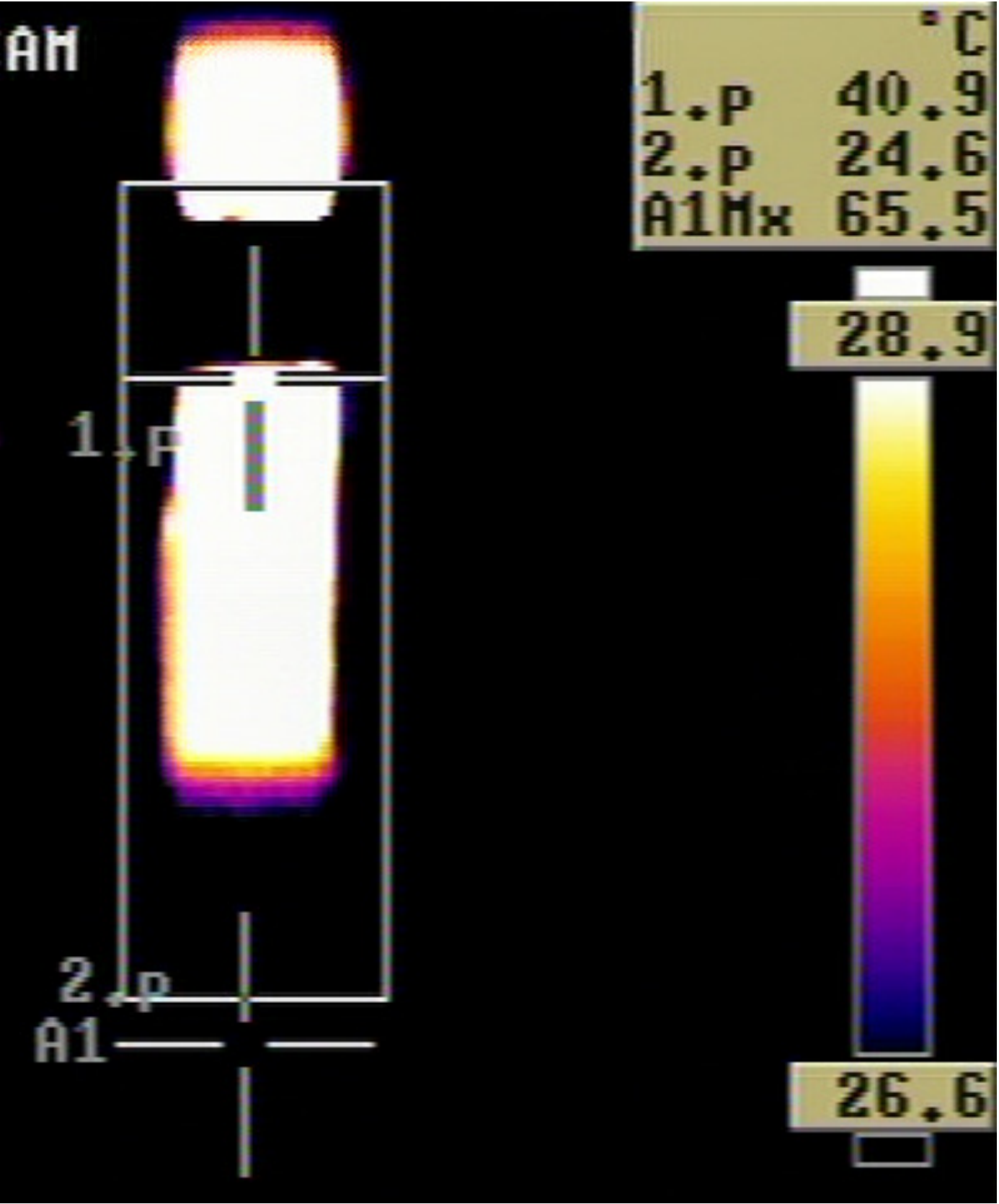}
 \hfill
\null
\caption{ Snapshots taken by the thermal camera. The first one shows the
initial state, then the quasi-adiabatic cooling is observable, then heat
dissipation appears due to plastic change, then the plastic change
reaches the whole thinner part of the sample, and finally failure
occurs.}
 \label{fbv}
\end{center}
}
\end{figure}

The same features can be observed in Fig.~\ref{fby}a, which presents the
time series of loading force and temperature of a similar experiment
performed on the same type of sample. Two loading--unloading cycles have
been carried out, the second with larger maximal stress than the first,
but both surely remained below the yield threshold. The third loading
was not terminated, in order to enter the regime of plastic change and
also to cause failure. Correspondingly, during the first two cycles,
temperature first drops a bit and then it returns. The third loading
also starts with cooling, until the plastic yield threshold is reached
(see the small transient in the stress curve), and subsequently plastic
dissipation starts to take the leading role, resulting in considerably
raising temperature. According to \re{@25278} with \re{fcg} [and
\re{fcx} below], the cooling part is linear in loading and the warming
part is quadratic/parabolic.

\setlength{\tffigw}{.39\columnwidth}
\begin{figure}[ht]
\parbox{77.5 mm}{%
\begin{center}
 \null
 \includegraphics[width=\tffigw]{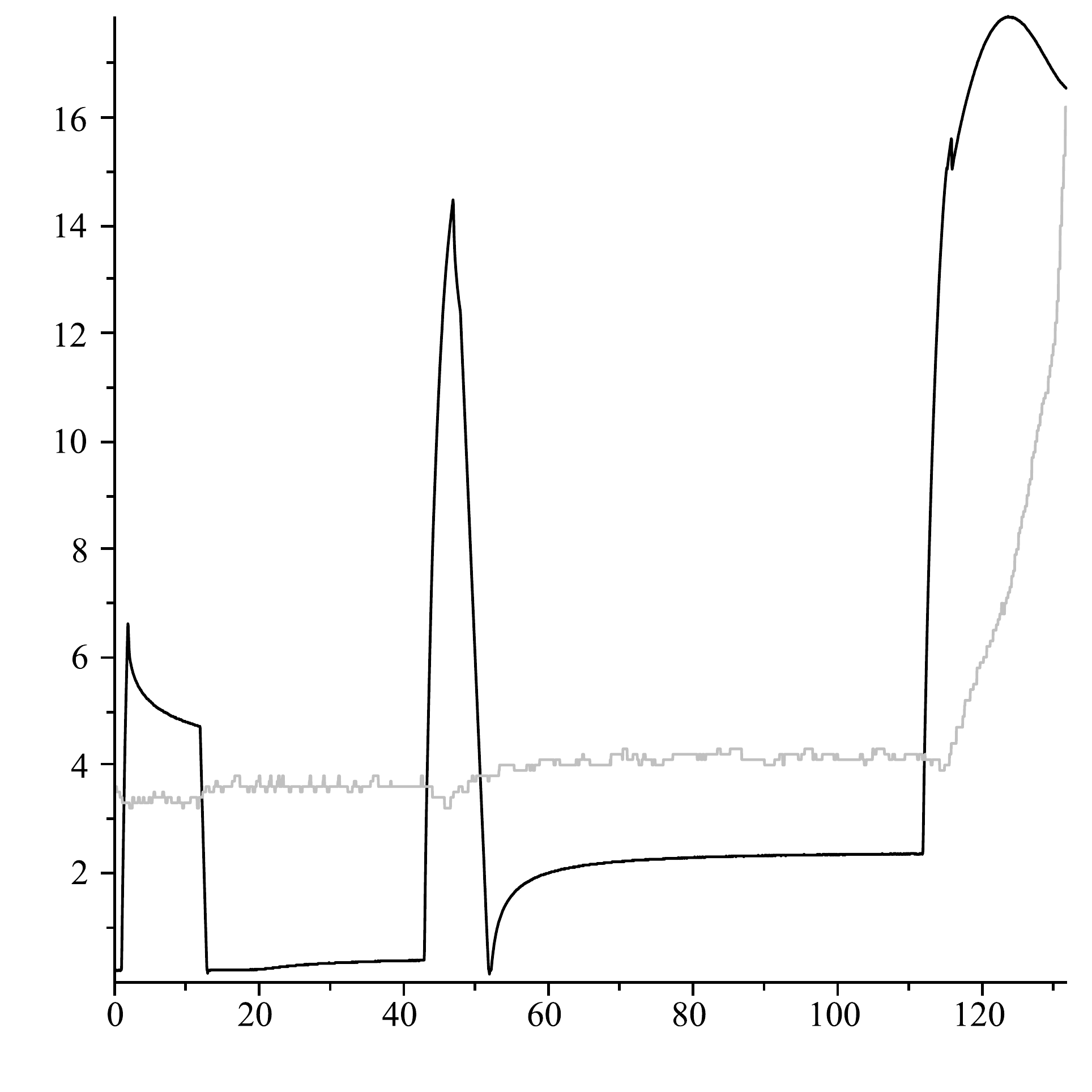}
 \hfill
 \includegraphics[width=\tffigw]{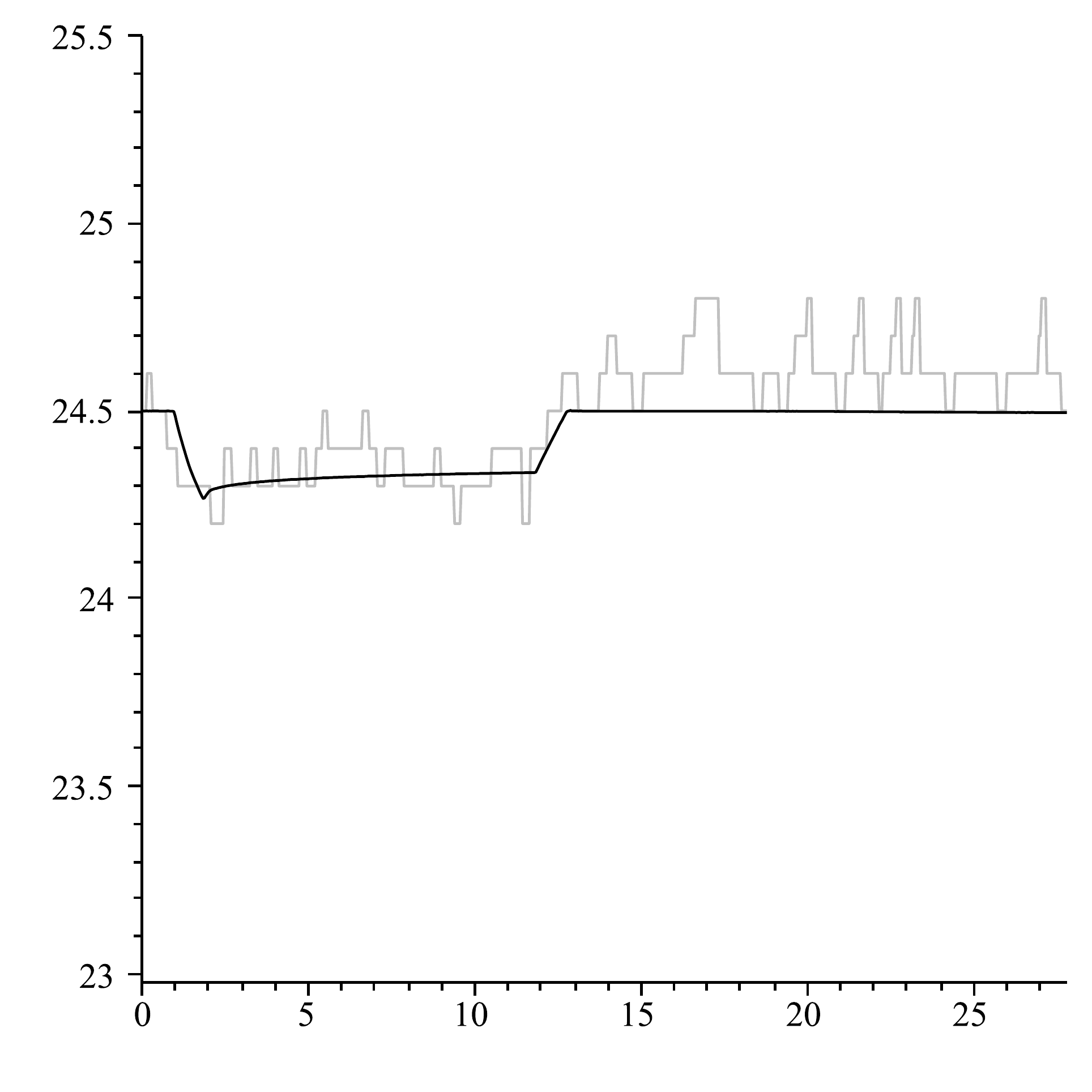}
 \null
\par\noindent
 \null
 \hfill
 \1 1 {a}
 \rule{.41\columnwidth}{0ex}%
 \1 1 {b}
 \hfill
 \null
\caption{(a) Measured temperature (grey line) and force (black line) as
a function of time, during two complete loading--unloading uniaxial
(strain-driven) stretching cycles followed by a loading until plastic
yield and failure. (b) Measured (grey line) and predicted (black line)
temperature as a function of time, during the first loading cycle.
(Force in units of 200 N, temperature in \m { ^\circ }C, time in s.)}
 \label{fby}
\end{center}
}
\end{figure}

In Fig.~\ref{fby}b, the first loading--unloading cycle is enlarged, and the
measured temperature is shown together with the prediction of the
theory, calculated via
 \eq{fcx}{
\Vrho \Vc \dot \VT = - \Valp \VT \dot\Vsig\para ,
 }
which follows from \re{@25278} with the absence of plastic change and
heat. Here, the literature constants \m { \Vrho = 1150 \:
\text{kg}/\text{m}^3 }, \m { \Vc = 1700 \: \text{J}/\1 1 { \text{kg}
\text{K} } } and \m { \Valp = 0.8 \cdot 10^{-6} \: 1/\text{K} } have
been used. The good agreement with measurement indicates that the theory
works well and the applied approximations are valid.

One can observe that, after a loading-unloading cycle, temperature does
not exactly return to the initial value but gets slightly raised. The
effect is larger for the second, larger loading cycle. Plasticity is
ruled out to be the cause of this dissipation. However, in fact there is
another aspect, rheology, also in action, inside such materials. In the
following section we show how rheology is required to be incorporated in
view of the experimental mechanical data, how nonequilibrium
thermodynamics provides us the required framework, how the data can be
fitted to determine the rheological coefficients, and that the presence
of rheology provides a correction not only for the mechanical side but
also contributes to the thermal changes as another source of
dissipation.

\section{The thermodynamical framework for introducing rheology}

Adding rheology to the theory is possible with the aid of internal
variables (dynamical degrees of freedom \cite{Ver97}). For the simpler
case of no thermal expansion and no plasticity considered, the
methodology has been given in \cite{AssFulVan14}. Here, we extend the
treatment for thermal expansion and plasticity included. Details
identical to the case of \cite{AssFulVan14} are only summarized here,
and here we focus on the differences.

According to the methodology, we assume the existence of a further
quantity, which is expected to be a symmetric tensor, corresponding to
that rheology manifests itself mainly in the mechanical behaviour, as an
additional source of stress -- such as an internal damping force. In
addition to the generalization of stress, we assume that specific
entropy, expressed as a function of \m { \VVD } and \m { \Ve }, now also
depends on the new variable \m { \VVxi }, via an additive quadratic term
that ensures that, in thermal equilibrium, entropy still gets maximal
(obeying the second law of thermodynamics). Namely, we have
 \eq{fch}{
\Vs \9 1 { \VVD, \Ve, \VVxi } = \Vs_{\text{previous}} \9 1 { \VVD, \Ve }
- \f{1}{2} \tr \0 1{ \VVxi^2 } .
 }
Taking its comoving time derivative, one can derive that entropy
production obtains, in addition to \re{fce}, two extra terms,
 \eq{@37874}{
\Vpi_{\Vs} & = \rtd\nabla \f{1}{\VT} \cdot \VVj_{\Ve} + \f{1}{\VT} \tr
\0 1 { \VVsig \VVZ }
 \non\lel{fcqq}
& \quad +
\f {1}{\VT} \tr \0 1{ \hat\VVsig \VVL\symm } - \Vrho \tr \0 1{ \VVxi^2 } ,
 }
where \m { \hat\VVsig } denotes the rheology-originated addition in
stress, added to the previous, elastic, stress. The Onsagerian way
to ensure the non-negativeness of this surplus entropy production
is to consider linear equations
 \eq{.13.57.}{
\hat\VVsig\dev & = \Vl_{11}\dev \9 1 {\VVL\symm}\dev +
\Vl_{12}\dev \2 1 {\mathord- \Vrho \VT \VVxi\dev} ,
 \lel{fcz}
\dot \VVxi{}\dev & = \Vl_{21}\dev \9 1 {\VVL\symm}\dev +
\Vl_{22}\dev \2 1 {\mathord- \Vrho \VT \VVxi\dev} ,
 \lel{fda}
\hat\VVsig\sph & = \Vl_{11}\sph \9 1 {\VVL\symm}\sph +
\Vl_{12}\sph \2 1 {\mathord- \Vrho \VT \VVxi\sph} ,
 \lel{.13.58.}
\dot \VVxi{}\sph & = \Vl_{21}\sph \9 1 {\VVL\symm}\sph +
\Vl_{22}\sph \2 1 {\mathord- \Vrho \VT \VVxi\sph} ,
 }
with positive definite matrices of coefficients
 \eq{@38881}{
\0 1{ \mat{ \Vl_{11}\dev & \Vl_{12}\dev \\
\rule{0em}{3.4ex} \Vl_{21}\dev & \Vl_{22}\dev } } ,
 \quad
\0 1{ \mat{ \Vl_{11}\sph & \Vl_{12}\sph \\
\rule{0em}{3.4ex} \Vl_{21}\sph & \Vl_{22}\sph } } .
 }
Note that one could introduce Onsagerian coupling to the plasticity
related term of \re{fcqq}, too. Nevertheless, for our current needs, the
present level of generality suffices. Namely, it successfully explains
all aspects of the experimental data that we consider here. On the other
side, without shifting the entropy -- and thus without the last term of
\re{fcqq} -- its term \m { \f {1}{\VT} \tr \0 1{ \hat\VVsig \VVL\symm }
} can be made positive definite via standard viscosity, \m {
\hat\VVsig\dev } being proportional to \m { \9 1 { \VVL\symm }\dev },
and \m { \hat\VVsig\sph } to \m { \9 1 { \VVL\symm }\sph }.

Hereafter, we discuss processes below the plastic yield threshold. In
addition, we neglect the tiny thermal expansion part in \re{@15991}
corresponding to the small temperature changes visible during loading
cycles like in Fig.~\ref{fby}. Then we are in the approximation
 \eq{@1599}{
\VVL\symm = \dot\VVD = \dot\VVeps .
 }
In parallel, with absolute temperature nearly constant, the coefficients
\m { \Vl } can safely be considered constant, and we can eliminate \m
{ \VVxi } from the set of equations \re{.13.57.}--\re{.13.58.}. The
elimination yields
 \eq{.13.59.}{
\VVsig\dev & + \VVtau\dev \dot\VVsig\dev
 \lel{fdi}\non
& = \VValp\dev \VVD\dev + \VVbet\dev \dot \VVD\dev + \VVgam\dev
\ddot \VVD\dev ,
 \lel{fdb}
\VVsig\sph & + \VVtau\sph \dot\VVsig\sph
 \lel{fdj}\non
& = \VValp\sph \VVD\sph + \VVbet\sph \dot \VVD\sph + \VVgam\sph
\ddot \VVD\sph
 }
for the total stress, where the new coefficients are straightforward
combinations of the former coefficients \m { \Vl }. These are two
separate linear rheological models with \m { \VVsig }, \m { \dot\VVsig
}, \m { \VVD }, \m { \dot\VVD }, \m { \ddot\VVD } terms. Therefore, the
classic Kelvin--Voigt, Maxwell and Jeffrey rheological models are covered
as special cases of these Kluitenberg--Verh\'as models (as named in
\cite{AssFulVan14}, see the explanation and further analysis there). An
important benefit of the thermodynamical derivation of these rheological
models is that, from the second law of thermodynamics, conditions follow
for the coefficients in \re{.13.59.}--\re{fdb} (which are the positive
definiteness criteria of the matrices \re{@38881} translated to these
coefficients, the simple derivation can be found in \cite{AssFulVan14}),
some of which are nontrivial and remarkable:
 \eq{@41262}{
\VVtau\dev & \ge 0 , & \VValp\dev & \equiv \VE\dev \ge 0 ,
 \lel{fdc}
\VVbet\dev & \ge \VVtau\dev \VValp\dev , & \VVgam\dev & \ge 0 ,
 \lel{fdd}
\VVtau\sph & \ge 0 , & \VValp\sph & \equiv \VE\sph \ge 0 ,
 \lel{fde}
\VVbet\sph & \ge \VVtau\sph \VValp\sph , & \VVgam\sph & \ge 0 .
 }

\section{Determining rheological constants from experimental data}

From experimental stress and strain values, one can determine the \m {
\VVtau }, \m { \VE_i } constants. The first step towards this is to
address the difference between \m{ \VVD } and \m{ \VVeps }. Since \m{
\VVD } may not start from zero at the initial time of an experiment --
some pre-stressing is usually needed to ensure the proper initial state
of the sample -- and as strain gauges (or other strain measuring
devices) may have some initial offset at initial time, one needs to
assume a difference between initial \m{ \VVD } and \m{ \VVeps }, which
leads to an offset \m { \VVdel }:
 \eq{.13.5.}{
\VVsig\dev & + \VVtau\dev \dot\VVsig\dev
 \lel{fdg}
& = \VVdel\dev + \VValp\dev \VVeps\dev + \VVbet\dev \dot \VVeps\dev
+ \VVgam\dev \ddot \VVeps\dev ,
 \non\lel{fdf}
\VVsig\sph & + \VVtau\sph \dot\VVsig\sph
 \lel{fdh}\non
& = \VVdel\sph + \VValp\sph \VVeps\sph_{} + \VVbet\sph \dot
\VVeps\sph_{} + \VVgam\sph \ddot \VVeps\sph_{} .
 }
The particular \m { \VVdel } values carry, therefore, no principal
information, as they are not material constants but just characterize
the experimental setup and circumstances.

The next is to realize the importance of having measurement data for \m
{ \Veps\orth } in uniaxial experiments. Namely, for uniaxial processes,
it is possible \cite{AssFulVan14} to derive from \re{.13.5.}--\re{fdf} a
relationship between \m { \Vsig\para } and \m { \Veps\para }. That
equation, however, contains derivatives up to the third derivative of \m
{ \Vsig\para } and fourth of \m { \Veps\para }, and includes all the
eight \m { \VVtau }, \m { \VE_i } constants. Both the high number of
constants and the extraction of such high derivatives from discretely
measured data -- that is burdened by error -- are considerable practical
difficulties. Moreover, even with precisely determined coefficients in
the relationship between \m { \Vsig\para } and \m { \Veps\para },
inverting these back to the \m { \VVtau }, \m { \VE_i } constants is
problematic because of the nonlinear formulae \cite{AssFulVan14}
connecting the two sets of coefficients. It is therefore crucial to have
reliable measurement of \m { \Veps\orth }, too, with which one can
calculate the deviatoric and spherical parts, and can then solve two
separate fitting problems for 4--4 coefficients only.

The equations \re{.13.5.} and \re{fdf} are linear in the unknown
parameters so a least-squares fitting seems feasible, with the data
values measured at the discretely many instants and the derivative
values also being derived from them. The third difficulty to face at is
that even the first and second derivatives are nontrivial to attribute
to a data series even if it has acceptably small errors: the errors get
seriously amplified in neighbouring differences and the fitted
parameters are unacceptably unreliable. The approach we have worked out
for treating data like in Fig.~\ref{fby}a is as follows.

\begin{figure}[ht]
\parbox{77.5 mm}{%
\begin{center}
 \includegraphics[width=.5\columnwidth]{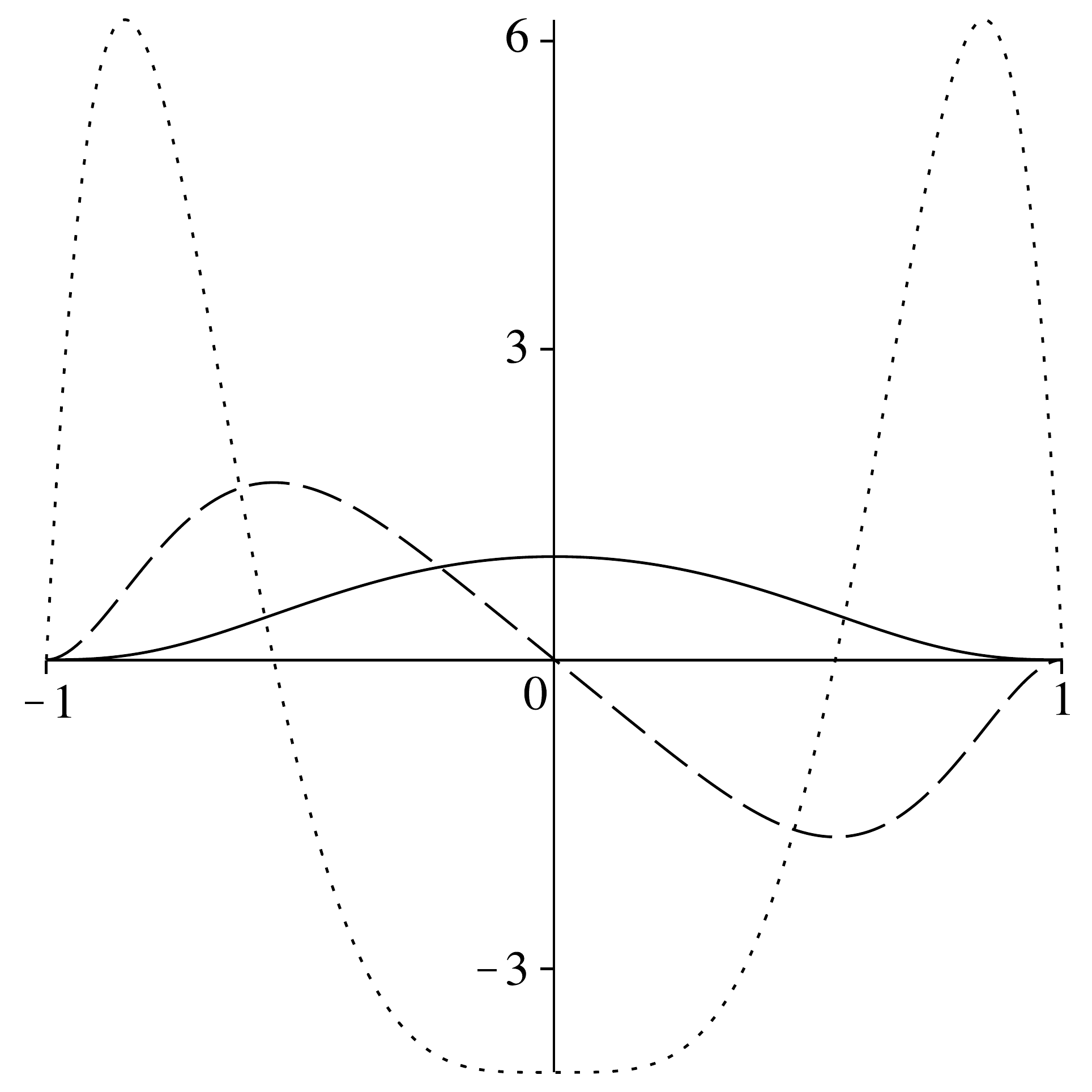}
\caption{The window function \re{@46080} (black line) and its first
(dashed line) and second (dotted line) derivatives. Each of these three
functions sample the interval \m { \1 2 { -1, 1 } } visibly uniformly
enough (not particularly favouring some parts of the interval over other
parts).}
 \label{fdm}
\end{center}
}
\end{figure}

We intend to perform some smoothing so, instead of treating an equation
like \re{.13.5.} or \re{fdf} directly, let us consider a time integral
of it. More precisely, we multiply the equation by a window function
centered around a time \m { t }, and integrate the product. The novelty
in doing this is that we choose such a window function \m { \Vw \1 1 { t
} } that is nonzero only in an interval \m { \1 2 { t_1, t_2 } } and it
becomes zero at the two interval endpoints such smoothly that even its
first and second derivatives tend to zero there. The advantage of this
is that, on the terms containing derivative, we can perform partial
integration with the benefit that the surface terms are zero, e.g.,
 \eq{@45376}{
\int_{t_1}^{t_2} \dot\Vsig\para \1 1 { t } \Vw\1 1 { t } \dd t
& = \2 2 { \Vsig\para \1 1 { t } \Vw \1 1 { t } }_{t_1}^{t_2} -
\int_{t_1}^{t_2} \Vsig\para \1 1 { t } \dot \Vw\1 1 { t } \dd t
 \non\lel{fdl}
& = - \int_{t_1}^{t_2} \Vsig\para \1 1 { t } \dot \Vw\1 1 { t } \dd t .
 }
This way, we can bring back all integrals of derivatives to integrals of
functions themselves (multiplied by the differentiated window function
that still behaves nicely). This approach is similar to the idea of the
so-called test functions in distribution theory. To ensure very fast
calculations, we have chosen \m { \Vw } to be a polynomial, and after
some experimenting, we have found the polynomial
 \eq{@46080}{
\Vp \1 1 { \Vu } = \1 1 { \Vu + 1 }^3 \1 1 { \Vu - 1 }^3 \1 1 { \Vu^2 +
1 } ,
 }
which apparently vanishes fast enough at the two endpoints of the
interval \m { \1 2 { -1, 1 } } (see Fig.~\ref{fdm}), to perform very
well. Then, for general intervals \m { \1 2 { t_1, t_2 } }, we have
rescaled its variable to obtain \m { \Vw \1 1 { t } }. One chooses at
least five different intervals and performs a least-squares fitting on
the integrated values.

The integral of any data series was calculated numerically, but here we
also wanted to improve the performance. The reason for this is that, in
many cases in practice, one has only a few dozens of data points at
hand. We need to consider at least five different intervals for
determining the five unknown constants, or, if we also want some error
bars for the determined constants then at least seven intervals are
preferable. Then the numerical integrals need to be fairly reliable
since their values will not differ too much, with the danger of
numerical unreliability.

To this end, we have modified the classic trapezoidal rule for numerical
integration. Namely, the trapezoidal rule approximates the integrand by
a linear function (piecewise). However, in our case, a product is to be
integrated, and the second derivative of our window function changes
unavoidably rapidly. This itself invalidates the (piecewise) linear
approximation considerably. The idea then is to approximate \textit{only
the data itself} by linear pieces, and to treat the window function (or
its appropriate derivative) in an exact manner. For example,
 \eq{@5376}{
& \int_{t_a}^{t_b} \Vsig\para \1 1 { t } \Vw\1 1 { t } \dd t
 \lel{fdk}\non
& \approx \int_{ta1}^{t_b} \9 2 { \Vsig\para \1 1 { t_a } + \f
{\Vsig\para \1 1 { t_b } - \Vsig\para \1 1 { t_a }}{t_b - t_a} \9 1 {t
- t_a} } \Vw\1 1 { t } \dd t ,
 }
the calculation of which going back to the integral of \m { \Vw \1 1 { t
} } and of \m { t \Vw \1 1 { t } }, which both we can determine exactly.

After these preparations, data behaving like those in Fig.~\ref{fby}a
proved to be treatable. For rheological constants, such parts of a time
series are informative where the data changes well enough so that its
second derivative also changes well enough. For demonstration, we show
here the fitting for just a small part of the first loading-unloading
cycle on Fig.~\ref{fby}a: the ending part of the loading and the
beginning part of the unloading. The abrupt change from loading to
unloading may seem dangerous to use but is actually most informative for
rheological parameters, which manifest themselves the best during heavy
changes.

The least-squares fitting provides us error bars and an \m { R^2 } value
can also be calculated to quantify the goodness of the fit, still, the
best is when one can also visualize the quality of the fit. To this end,
solving the simple numerical scheme
 \eq{@49524}{
\Vsig\sph_n & + \VVtau\sph \f {\Vsig\sph_{n\vphantom{1}} -
\Vsig\sph_{n-1}}{\Delta t}
 \non\lel{fdn}\non
& = \Vdel\sph + \VValp\sph \Veps\sph_n
 + \VVbet\sph \f{\Veps\sph_{n\vphantom{1}} - \Veps\sph_{n-1}}{\Delta t}
 \lel{fdp}
& + \VVgam\sph \f{\Veps\sph_{n+1} - 2 \Veps\sph_{n\vphantom{1}} +
\Veps\sph_{n-1}}{\Delta t^2}
 }
for \m { \Vsig\sph_n }, we predicted \m { \Vsig\sph } values from
experimental \m { \Veps\sph } values and an initial \m { \Vsig\sph }
value (and treated the deviatoric counterpart analogously). An advantage
of this scheme is that it is reliably applicable even if some of the
fitted coefficients are small, or are known to be zero a priori.

As mentioned, we present here the result for just a small part of the
process, in order to exhibit the quality of the established numerical
approach. Thirty data points are used, and integration is done on seven
intervals. This way, each interval consists of only six points.
(Neighbouring intervals have an overlap of two points.) Apparently, six
points means a rude discretization for an integral on an interval so
such a situation is a good test of the numerical method. The obtained
fitted coefficients and the predictions based on them can be seen in
Table~\ref{fdr} and Fig.~\ref{fdq}.

\nc\yhely{\vphantom{\m { |^|_+ }}}
\begin{table}[ht]
\parbox{77.5 mm}{%
\begin{center}
 \begin{tabular}{|l|c|c|}  \hline
\multicolumn{1}{|c|}{material}  &  fitted  &  standard  \\
\multicolumn{1}{|c|}{parameter}  &  value  &  error  \\ \hline \hline
\m{ \VVtau\dev \1 2 { \text{s} } }  &  0.3600  &  \m{\pm 0.0659}
 \yhely \\ \hline
\m{ \VValp\dev \1 2 { \text{GPa} } }  &  0.8612  &  \m{\pm 0.0556}
 \yhely \\ \hline
\m{ \VVbet\dev \1 2 { \text{GPa}\cdot \text{s} } }  &  0.4724  &
\m{\pm 0.0686}
 \yhely \\ \hline
\m{ \VVgam\dev \1 2 { \text{GPa}\cdot \text{s}^2 } }  &  0.0029  &
\m{\pm 0.0010}
 \yhely \\ \hline
 \hline
\m{ \VVtau\sph \1 2 { \text{s} } }  &  0.2329  &  \m{\pm 0.0904}
 \yhely \\ \hline
\m{ \VValp\sph \1 2 { \text{GPa} } }  &  4.5708  &  \m{\pm 1.0392}
 \yhely \\ \hline
\m{ \VVbet\sph \1 2 { \text{GPa}\cdot \text{s} } }  &  1.8566  &
\m{\pm 0.4401}
 \yhely \\ \hline
\m{ \VVgam\sph \1 2 { \text{GPa}\cdot \text{s}^2 } }  &  0.0013  &
\m{\pm 0.0220}
 \yhely \\ \hline
 \end{tabular}
\caption{
Table of fitted rheological coefficients for the deviatoric and
spherical parts}
 \label{fdr}
\end{center}
}
\end{table}

\setlength{\tffigw}{.42\columnwidth}
\begin{figure}[ht]%
\parbox{77.5 mm}{
\begin{center}
 \includegraphics[width=\tffigw]{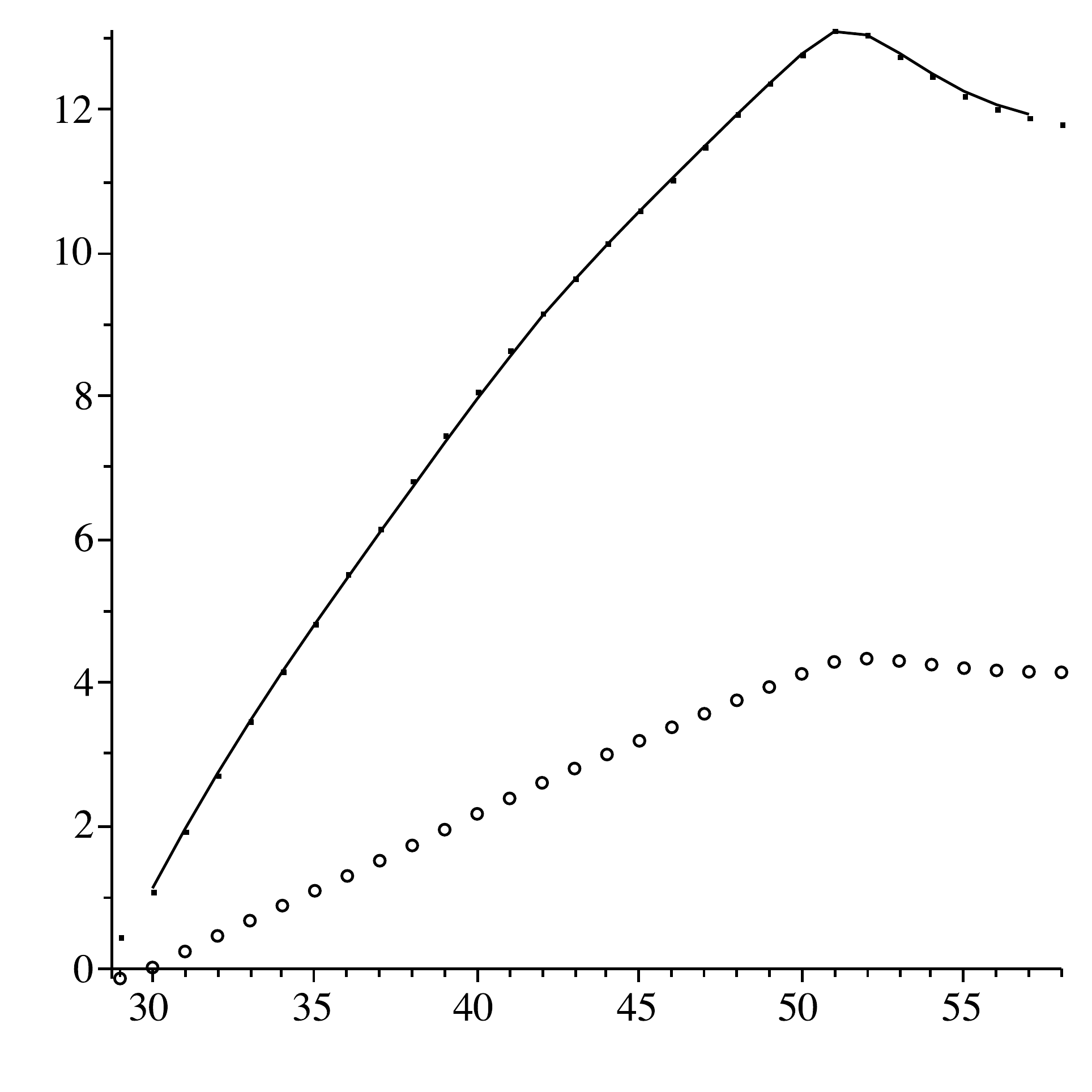}
 \hfill
 \includegraphics[width=\tffigw]{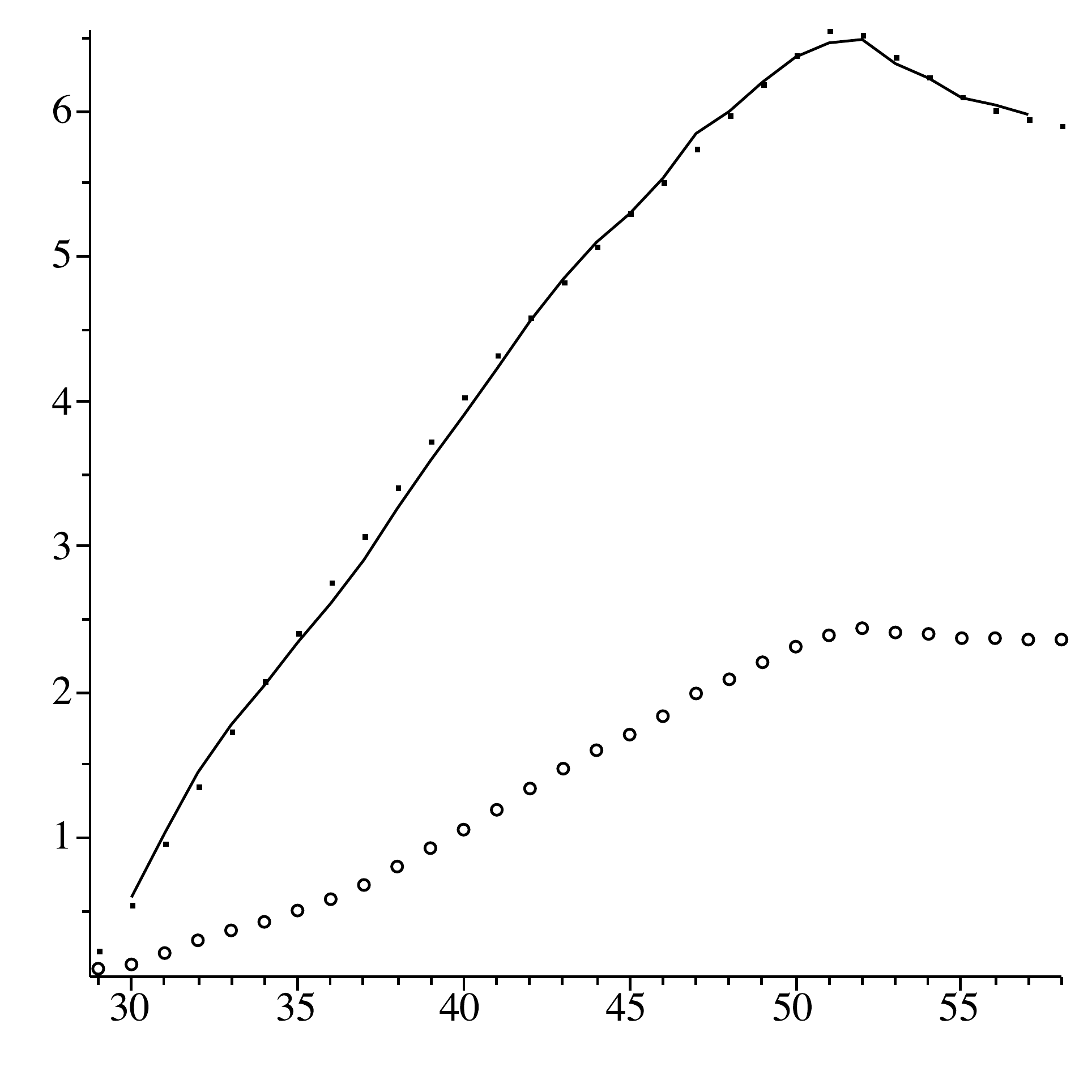}
\par\noindent
 \null
 \hfill
 \1 1 {a}
 \rule{.38\columnwidth}{0ex}%
 \1 1 {b}
 \hfill
 \null
\caption{Comparison of the theoretical prediction based on the values of
Table~\ref{fdr} (solid line) to the experimental stress data (points).
For information, the measured strain values (circles) are also displayed
(rescaled appropriately). Horizontal axes: time in s, vertical axes:
stress in MPa. (a) Deviatoric case. (b) Spherical case.}
 \label{fdq}
\end{center}
}
\end{figure}

Various further checks can be made for the determined rheological
coefficients. First, it is reassuring to find that the thermodynamical
constraints \re{@41262}--\re{fde} are fulfilled. Second, the elastic
constants \m { \VValp\dev }, \m { \VValp\sph } enable to calculate
Poisson's ratio and Young's modulus (see Appendix~A of
\cite{AssFulVan14}). The resulting former value (0.37) coincides nicely
with the literature data (0.38). Young's modulus (1.2 GPa) comes out
somewhat below the typical literature range (1.9 GPa \m { \sim } 3.3
GPa) which, however, depends heavily on humidity (that's why the
literature range is so wide). In parallel, one must bear in mind that
customary measurements of Young's modulus are performed at finite
loading speeds where the steepness of the longitudinal stress--strain
curve is seriously influenced by rheology. For example, even for the
simplest rheological situation, a Kelvin--Voigt model in the deviatoric
part and a Hooke model in the spherical part, the uniaxial equation
contains not only \m { \Vsig\para } and \m { \Veps\para } but also their
first time derivative, too (\cite{AssFulVan14}, Appendix~A). The ratio
of these derivatives, \m { \dot\Vsig\para / \dot\Veps\para }, dominates
the slope of the longitudinal stress--strain curve at the beginning of
the loading and, for not too slow loading, the later part as well. In
other words, the ratio of the coefficients of these first derivatives
(in systematic notation, \m { {\VVbet}\para / {\VVtau_1}\para }) plays the
role of a dynamical Young's modulus. Now, for the rheological
coefficients found by the above fit, this dynamical Young's modulus
turns out to be 1.47 GPa, 24\% higher than the static one. We remark
that, as a consequence of the thermodynamical inequalities
\re{@41262}--\re{fde}, the dynamical Young's modulus is always larger
than the static one \cite{AssFulVan14}. For more complicated rheologies
-- like the one we have here -- even higher order dynamical Young's
moduli are present (ratios of coefficients of higher derivatives). The
second order dynamical Young's modulus, for example, is already 1.84
GPa.

It is a generally valid warning from the rheology of solids that
customary ways of extracting Young's modulus from longitudinal
stress--strain curves may result in erroneously high values whenever
rheology is neglected. Notably, it is not only plastics \cite{KocHor13}
and similar materials where one may encounter rheology in solids: as an
important application, the mechanical description of rocks also requires
the full rheological model, as exploited in the ASR (Anelastic Strain
Recovery) method for determining 3D in situ stress (see
\cite{MatTak93,Mat08,LinEta10}).

\section{Discussion}

The presented theory is rich in the sense that it grasps the elastic,
thermal expansion, thermal conduction, rheological and plastic aspects
of solids. On the other side, for each of these aspects, the simplest
available kinematic and constitutive choices have been made here, in
order to make the comparison with experimental data easy. These specific
choices have been found to be in conform with the experiments that we
have performed for illustration. The framework is nevertheless general
enough to describe much more general processes and material behaviours
as well, such as large deformation processes, material anisotropy,
nonlinear elasticity, elaborated plastic effects, and all with
nonconstant elasticity, thermal expansion, rheological and other
coefficients.

As for feasible future generalizations of the theory, first, one can
mention allowing the Onsagerian coupling between the plastic and the
rheological side as well. Second, non-Fourier heat conduction can also
be incorporated, unifying the internal variable approach \cite{VanFul12}
for rigid heat conductors with the thermomechanical side described here.
Damage and failure \cite{VanVas14,DeaEta12} are more distant but also
reasonable candidates to include via the same thermodynamical
methodology, along the lines of the works \cite{Van96,VanVas01}.
Temperature dependent complicated rheological/viscoelastic situations in
the finite deformation regime \cite{PalVar10} mean further challenges.

On the experimental side, our intention here was to present examples for
the involved various aspects. These aspects, such as the Joule-Thomson
effect and the Kluitenberg--Verh\'as rheology, have been found to be in
good quantitative agreement with the theory. In the future, with higher
precision and higher reliability measurements of the kind performed
here, a full quantitative agreement can be achieved. For example,
already the present data allows us to approximately identify the plastic
yield stress value to be around 100 MPa (the stress level where the
transient takes place in Fig.~\ref{fby}a and temperature starts showing
plastic dissipation), and the ratio of the steepnesses before and after
the plastic yield leads to an approximate prediction of the plastic rate
coefficient [\m { \Vgam \approx 0.17 } in \re{fcs}] but a more elaborate
approach is desirable. If the experimental setup is able to provide fast
enough loading and unloading --needed for studying rheology precisely --
while all strain, force and temperature values are measured with high
precision and reliability then, after fitting all the material constants
involved, the numerical scheme explained in Sect.~\ref{fdv} (and
extended to also include thermal expansion and plasticity) should be
able to reproduce the whole time series precisely.

In addition to discussing the numerical scheme for predicting the time
dependence of quantities, we have presented a method for fitting the
rheological coefficients, which succeeded in this task satisfactorily.
We have emphasized the importance of having transversal strain to be
also measured, since this reduces the labour of fitting significantly,
enabling to divide the problem into two separate, easily treatable,
halves (deviatoric and spherical), while the composite uniaxial
eight-parameter situation may be found practically intractable.

Besides monitoring the mechanical quantities, measuring temperature
provides valuable additional information about the process of solid
samples. The elastic, rheological and plastic aspects make a clear
footprint on the temperature time series as well. Actually, because of
the inherent intertwining of thermal and mechanical aspects, the set of
equations is closed only when the temperature variable is a full member
of the set of variables. The thermodynamics-based approach enlights the
importance of temperature for processes that are traditionally treated
as mechanical-only.

The analysis of the experimental data has clearly shown the relevance of
rheology for plastics, and the situation is known to be the same for
seemingly more "solid" solids like rocks \cite{MatTak93,Mat08,LinEta10},
which is the base of the above-mentioned ASR method for determining 3D
in situ stress. What seems nonlinear elasticity in the stress--strain
plane may well be rheology. However, another important remark to be made
is that delays in an experimental equipment (implementation of loading,
controlling, measuring devices etc.) should not confused with
rheological delays acting within the sample. Much experimental care is
needed, in parallel to the proper theoretical interpretation, to
identify the presence of true rheology in solids. The applicational
consequences of reliable rheological information are far-reaching (the
long-time behaviour of underground tunnels, safety aspects of structural
materials etc.). The interplay between thermal, elastic and rheological
effects is also remarkably important. The thermodynamics-based
description of solids provides a reliable theoretical framework for all
these phenomena.

\section*{Acknowledgments}

The work was supported by the grants OTKA K81161
 and K116197.
Financial support from the Bolyai Scholarship of the Hungarian Academy
of Sciences is gratefully acknowledged by Attila Csat\'ar PhD.

\appendix

\section{Technical details of the measurement}

\begin{figure}[ht]
\parbox{77.5 mm}{%
\begin{center}
 \includegraphics[width=.42\columnwidth]{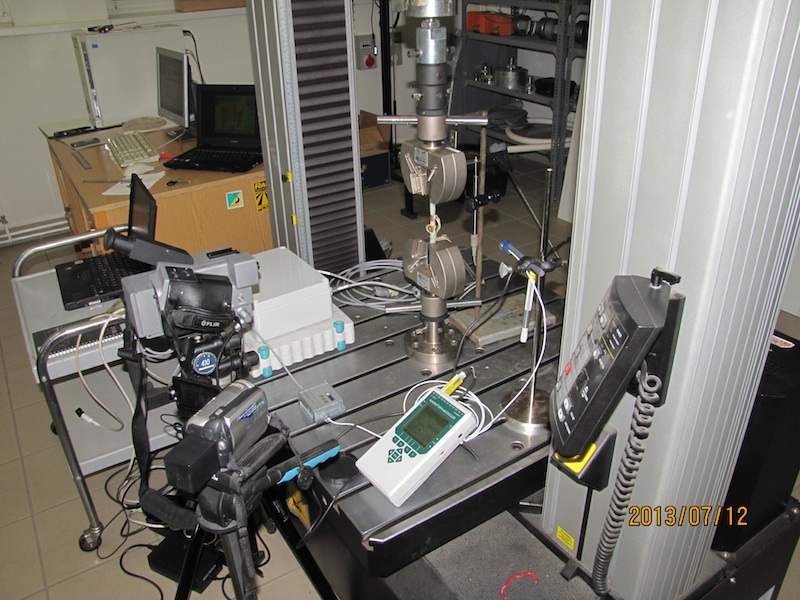}
\hfill
 \includegraphics[width=.42\columnwidth]{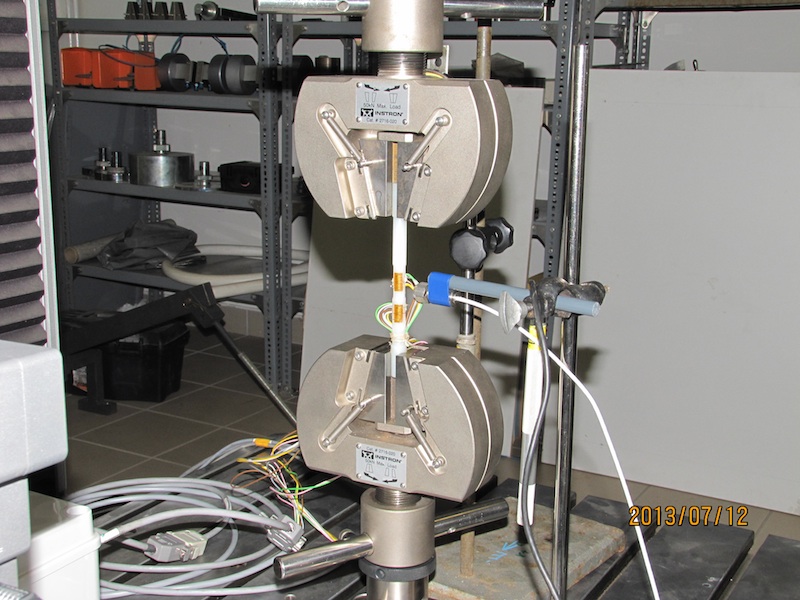}
\caption{The measuring arrangement}
 \label{fdt}
\end{center}
}
\end{figure}

The measurements have been carried out with an Instron 5581 universal
material tester device, at the Hungarian Institute of Agricultural
Engineering. The arrangement can be seen in Fig.~\ref{fdt}.

\begin{figure}[ht]
\parbox{77.5 mm}{%
\begin{center}
 \includegraphics[width=.42\columnwidth]{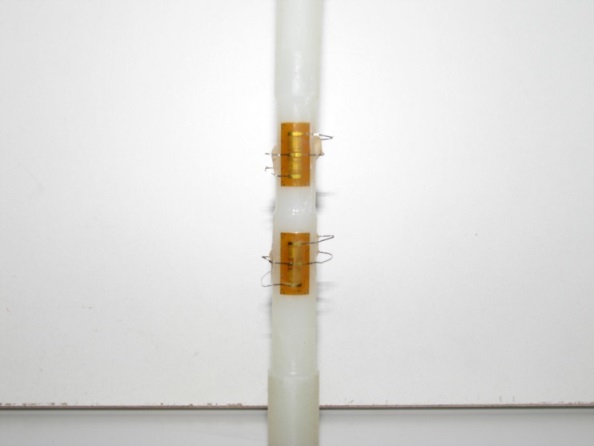}
\caption{Strain gauge used for the measurement}
 \label{fdu}
\end{center}
}
\end{figure}

To track the longitudinal and lateral size changes of the tested sample,
a tensiometric measuring device was used. HBM 3/350 XY11 type strain
gauge was mounted onto the specimen (Fig.~\ref{fdu}). The resistance of
the gauge is \m{R = 350 \Omega}, and the gauge factor is \m{k=1.98}
(1\%, according to the technical parameters provided by the supplier.)

Half bridge was used to test the specimen. The active gauge was the one
mounted on the specimen, while the other gauge of the bridge was mounted
on an unloaded metal plate. During the measurements, the half bridge was
connected to an SR-55 radio frequency module of a Spider-8 amplifier.

A ThermaCAM PM695 type real-time thermal camera was used for temperature
measurement and imaging. It monitored the area of the specimen indicated
by dashed lines in Fig.~\ref{fbu}. Besides the average temperature of
the area and the maximal temperature value within the area, the
temperature value at two fixed spots were also registered. In parallel,
temperature at some additional spots was monitored via infrared
temperature sensor (supplier: Optrics GmbH, type: OPTCTLT10FCB3).

Standard specimens were used. After manufacturing, the specimens were
relieved from load, and were calibrated after the gauges were mounted.
The geometric dimensions of the specimen are displayed in
Fig.~\ref{fbu}.

\bibliography{godollo-ft}
\bibliographystyle{unsrt}

\end{document}